\pdfoutput=1
\documentclass[10pt,journal]{IEEEtran}

\usepackage{amsmath,amsfonts}
\usepackage{algorithm}
\usepackage{algpseudocode}  
\usepackage{array}
\usepackage[caption=false,font=normalsize,labelfont=sf,textfont=sf]{subfig}
\usepackage{textcomp}
\usepackage{stfloats}
\usepackage{url}
\usepackage{verbatim}
\usepackage{graphicx}
\usepackage{tikz}
\usetikzlibrary{positioning,calc,arrows}
\usepackage{cite}
\usepackage{braket}
\usepackage{mathrsfs}
\usepackage{amssymb, amsthm}
\usepackage{multirow}
\usepackage{hhline}
\usepackage{makecell}
\usepackage{color}
\usepackage{mathtools} 
\usepackage{enumitem}
\usepackage{booktabs}
\theoremstyle{definition}

\begin{document}
\title{Offline Channel-Independent QAOA Angles for RIS Power Aggregation: Unit-Circle Phase Dictionaries and Infinite-Size Spin-Glass Limits}
\author{Burhan Gulbahar
\thanks{Burhan Gulbahar is with the Department of Electrical and Electronics Engineering, Ya{\c{s}}ar University, 35100 Izmir, Turkey (e-mail: burhan.gulbahar@yasar.edu.tr).}
\thanks{This work has been submitted to the IEEE for possible publication. Copyright may be transferred without notice, after which this version may no longer be accessible.}
}
\maketitle
\begin{abstract}
Reconfigurable intelligent surfaces (RIS)  maximize received power  by setting per-element phases. Discrete-phase optimization is NP-hard in the worst case, while the quantum approximate optimization algorithm (QAOA) applied to RIS faces limited phase alphabets, either per-problem angle optimization or uncharacterized training cost exposed to barren plateaus, and no scalable performance benchmark. We introduce a $2^{M}$-phase  $\theta$ dictionary for optimizing power $\|\mathbf{A} \, e^{j\theta}\|^{2}$  having $K \times N$ channel matrix $\mathbf{A}$ and  QAOA angle offline optimization with instance and size-independent infinite-size limit of the mixed-$q$ Gaussian ensemble of Basso et al. Our design bounds the spin-Hamiltonian interaction order to at most quartic for any $M$, and the deployed order-2 reduction lies below the even-$q\!\ge\!4$ regime in which constant-level QAOA limitations are proved. We perform analytical, state-vector, matrix-product-state and Pauli-path-simulation numerical studies for   $N=K \leq 100$ and QAOA depth $p=9$, verifying offline angle transfer to  Rayleigh, Rician/line-of-sight, cascaded double-fading and spatially-correlated RIS channels at $N\!\in\!\{5,12\}$. We observe performance reaching a near-optimal multi-start  single-flip local-search reference for $N\!\le\!16$ under order-2 modeling with $2^{5}{=}32$-phase dictionary while the order-4 model shows a performance ceiling below the classical reference. The approach suggests a route to near-optimal large-$N$ performance on future fault-tolerant (FTQ) quantum computers, which enable the higher-depth QAOA circuits.
\end{abstract}
\begin{IEEEkeywords}
QAOA, reconfigurable intelligent surfaces, phase-only optimization, power aggregation, higher-order binary optimization, HUBO, Ising formulation.
\end{IEEEkeywords}

\section{Introduction}
Reconfigurable intelligent surfaces (RIS) enable programmable control of electromagnetic propagation through per-element phase adjustments~\cite{ElMossallamy2020}. A central phase-only design task is to maximize the aggregate received power generated by a surface configuration, which leads naturally to objectives of the form $\max_{\Omega}\|A e^{j\Omega}\|_2^2$, where $A$ is an effective RIS channel matrix and $\Omega$ contains the RIS phases~\cite{xiong2025das}. This formulation appears in structured physics-based models~\cite{ElMossallamy2020}, in cascaded system-level models~\cite{Zhang2023StatCSI}, and in multi-user (MU) power-aggregation settings~\cite{wuzhang2019sdr,xiong2025das}. However, unit-modulus or discrete phase constraints make the optimization a highly nonconvex, NP-hard combinatorial problem~\cite{zhang2006complex}.
 
This worst-case hardness has motivated interest in quantum optimizers, notably the quantum approximate optimization algorithm (QAOA), for wireless design~\cite{krikidis2026qow}. Quantum-annealing (QA) formulations map low-bit RIS phase selection onto Ising hardware, radar-cross-section (RCS) engineering on D-Wave~\cite{ross2022engineering,lim2024taprls}, hybrid loops~\cite{ross2023nemo,qaem4eng2025}, cell-free modeling~\cite{ohyama2022cellfree}, and 1-bit index modulation~\cite{ris1bit2025}, while gate-based QAOA targets spin-glass~\cite{colella2024multipath}, physics-informed RCS~\cite{pasquale2026piqaoa}, simultaneously transmitting and reflecting (STAR) RIS variants~\cite{pham2026starjoint,lim2026starcoupled,paul2026qaqgnf,sharma2026pqgd}, and quantum graph neural networks (QGNNs)~\cite{qgnn2026}. Across these, the phase alphabet is mostly one or two bits per element, larger surfaces are reached only through tile/block decompositions or outer classical loops, higher-order interactions are truncated to pairwise couplings, and no constant-level performance benchmark is supplied. 

In this work, we present a codebook-level reformulation lifting these restrictions: an exact $L=2^{M}$-phase nonuniform dictionary whose design bounds the induced spin-Hamiltonian order at four for any $M$ while a direct binary phase encoding instead grows the order as $2 \,M$. We reduce to order-2 regime by design where  channel-independent QAOA angle set is optimized with an instance-free Basso--Gamarnik--Mei--Zhou (BGMZ) infinite-size energy-density $V_p$ performance benchmark. The overlap-gap property (OGP) is proved to obstruct constant-level QAOA for pure even-$q\!\ge\!4$ spin models~\cite{BassoFOCS22}; the order-2 sector lies outside that theorem's scope. We synthesize the hard-unit-circle dictionary, write the  objective as a quadratic/higher-order binary optimization (QUBO/HUBO) problem, connect it to realistic links through a cascaded reduction with a common stochastic channel ensemble and position  QAOA as a benchmark. We validate the offline angle transfer on physically-grounded RIS channels, Rician/line-of-sight (LOS), cascaded double-fading  and spatially-correlated fading, at $N\!\in\!\{5,12\}$. Since our design with deeper, densely-connected circuits at large $N$ exceeds the reach of noisy intermediate-scale (NISQ) hardware, we evaluate it entirely in classical simulation and position it for the fault-tolerant (FTQ) regime rather than near-term devices.
\subsection{Contributions}
The main contributions are as follows:
\begin{enumerate}[leftmargin=1.15em]
\item \textit{Offline, instance and size-independent BGMZ angle design.} We optimize power aggregation  $\|\mathbf{A} \, e^{j\theta}\|^{2}$  having $K \times N$ RIS channel matrix $\mathbf{A}$ with $p$-layer QAOA. Conventional QAOA is either  \emph{online} by re-optimizing angles per instance as a deployment bottleneck for RIS with changing  channel $\mathbf{A}$ or \emph{train once} by reusing the angles but uncharacterized scaling cost of training vs. array size and exposed to barren plateaus   \cite{colella2025barren}. Our angles are instead fixed \emph{offline} by classically optimizing deterministic infinite-size $V_p$ at a cost exponential in  $p$ ($M_p = 4^p$) but independent of instance and problem sizes \cite{BassoFOCS22,Basso2021}. To
our knowledge this is the first instance  and problem-size independent QAOA angle design for RIS phases, valid at every $N$ since $V_p$ is an $n\!\to\!\infty$ object  as an instance-free mean-energy benchmark.
\item \textit{Hard-unit-circle dictionary with design-and-snap  formation.} An exact discrete phase dictionary is synthesized from shared binary sign patterns through linear and quadratic templates under exact
unit-circle equalities, casting RIS power aggregation as a binary--quadratic
optimization over sign matrices. The induced Hamiltonian has order $q=2 \, d$
($d{=}1$: order-2  surrogate, exact only for the degenerate four-phase unit-circle dictionary; $d{=}2$: exact dense order-4 HUBO). Since the templates are at most quadratic, this order is \emph{bounded by four for any $L=2^{M}$ phase resolution} by decoupling the resolution from the order and  keeping constant-level QAOA applicable at high resolution.  
\item \textit{Order-2 simulations and large-$N$ 
estimate.}  Simulations are performed for $ K = N \leq 100$ and  $p\le 9$ with exact state-vector (SV), matrix-product-state (MPS)  or Pauli-path-simulation (PPS) methods. The best QAOA sample reaches the performance of a multi-start single-flip local-search reference for $N \leq 16$  with monotonically rising sample mean \cite{xiong2025das}. Angles further transfer to Rician/LOS, cascaded double-fading, and spatially-correlated  channels at $N\!\in\!\{5,12\}$, re-measuring only the scalar local-field $c_2$.
\item \textit{Performance of the order-4 HUBO model.} We observe depth-independent  best sample plateaus below the local-search reference with finite-size concentration consistent with exponential sampling obstruction for pure even-$q\!\ge\!4$ spin models, motivating the order reduction~\cite{BassoFOCS22,Basso2021}.
\end{enumerate}
\begin{table}[!t]
\centering\footnotesize
\caption{Symbols and Meanings}
\label{tab:symbols}
\setlength{\tabcolsep}{2pt}
\renewcommand{\arraystretch}{1.05}
\begin{tabular}{|p{1.15cm}|p{3cm}|p{1.05cm}|p{3.1cm}|}
\hline
\textbf{Sym.} & \textbf{Meaning} & \textbf{Sym.} & \textbf{Meaning}\\
\hline\hline
$N$ & RIS elements & $M$ & bits/element, $L{=}2^{M}$ levels\\\hline
$K$ & observation ports & $A$ & channel matrix ($K\!\times\!N$)\\\hline
$n{=}NM$ & total spins (qubits) & $p$ & QAOA depth\\\hline
$\mathcal D$ & sign dict.\ ($L\!\times\!M$) & $T$ & sign variable ($N\!\times\!M$)\\\hline
$(\gamma,\beta)$ & QAOA angle vectors & $q{=}2d$ & order ($d$: template deg.)\\\hline
$C,\langle C\rangle$ & power cost; its mean & $C_0$ & offset $\|A\|_F^2$\\\hline
$C_{\mathrm{opt}}$ & order-2 ref. opt. & $r_b$ & best $C/C_{\mathrm{opt}}$\\\hline
$r_m$ & mean $\langle C\rangle/C_{\mathrm{opt}}$ & $r_m^{n}$ & offset-subtracted $r_m$\\\hline
$d_1$ & $p{=}1$ density & $\rho$ & $d_1^{\mathrm{RIS}}/d_1^{\mathrm{Gauss}}$\\\hline
$r_{\mathrm{eff}}$ & eff.\ rank of $(J_{uv})$ & $V_p$ & BGMZ depth-$p$ value\\\hline
$c_q$ & coeff.\ $\sqrt{(q{-}1)!D_q\overline V_q}$ & $\overline V_q$,\ $D_q$ & mean-sq. coeff.; deg./spin\\\hline
$\chi,\epsilon$,\ $w$ & MPS bond/trunc.; PPS wt. & $P_\alpha$ & $P(C\!\ge\!\alpha \, C_{\mathrm{opt}})$\\\hline
\end{tabular}
\end{table}
\begin{table}[!t]
\caption{Quantum-assisted RIS phase optimizers}
\label{tab:relwork-quantum}
\centering\footnotesize\setlength{\tabcolsep}{1.5pt}
\renewcommand{\arraystretch}{1.05}
\begin{tabular}{|p{1.25cm}|p{0.85cm}|p{1.35cm}|p{1.7cm}|p{0.7cm}|p{1.0cm}|p{1.15cm}|}
\hline
\textbf{Ref.} & \textbf{Alg.} & \textbf{Link} & \textbf{$N$} & \multicolumn{2}{c|}{\textbf{Obj.\ $\|\mathbf{A}e^{j\theta}\|^{2}$}} & \textbf{Opt.} \\
\cline{5-6}
 &  &  &  & \textbf{$\mathbf{A}$} & \textbf{$\theta$} &  \\
\hline\hline
\multicolumn{7}{|c|}{\textbf{Quantum annealing and other quantum optimizers}} \\
\hline
\cite{ross2022engineering,lim2024taprls} & QA & RCS & HW: $N\!\le\!177$ & $K\!\times\!N$ & $\le\!2^{2}$ & online \\\hline
\cite{ross2023nemo} & QA & SISO & sim: $N = 100$ & $1\!\times\!N$ & $\le\!2^{2}$ & online \\\hline
\cite{ohyama2022cellfree} & QA & MU/MIMO & HW: $N\!=\!16$ & --- & $72$-lvl & online \\\hline
\cite{ris1bit2025} & QA & SISO/IM & HW: $N\!\le\!100$ & $1\!\times\!N$ & $2^{1}$ & online \\\hline
\cite{sharma2026pqgd} & QGD & STAR/FD & sim: $N\!\le\!250$ & --- & cont. & online \\\hline
\cite{qgnn2026} & QGNN & MU & HW: $N\!=\!40$ & --- & discrete & online \\\hline
\hline
\multicolumn{7}{|c|}{\textbf{QAOA-based optimizers}} \\
\hline
\cite{colella2024multipath} & QAOA & MU & sim: $N\!=\!9$ & $K\!\times\!N$ & $2^{1}$ & train-once \\\hline
\cite{pasquale2026piqaoa} & QAOA & RCS & sim: $N\!=\!25$ & $1\!\times\!N$ & $2^{1}$ & online \\\hline
\cite{pham2026starjoint} & QAOA & STAR/MU & sim: $N\!\le\!18$ & $K\!\times\!N$ & $\le\!2^{2}$ & online \\\hline
\cite{lim2026starcoupled} & QAOA & STAR & sim: $N\!=\!64$ & --- & $2^{2}$ & online \\\hline
\cite{paul2026qaqgnf} & QAOA & STAR/NF & sim: $N\!=\!192$ & --- & $2^{2}$+cont. & online \\\hline\hline
\textbf{This work} & \textbf{QAOA} & \textbf{MISO/MU} & \textbf{sim: $N\!\le\!100$} & \textbf{$K\!\times\!N$} & \textbf{$2^{M}$} & \textbf{offline} \\\hline
\end{tabular}
\end{table}
\subsection{Related Works}
\label{subsec:relwork}
  
Table~\ref{tab:symbols} shows the symbols in the article. Classical methods report $N\!\sim\!10^{2}$--$10^{3}$ in simulation (sim.) and quantum-assisted ones $N\!\sim\!10^{1}$--$10^{2}$ on hardware (HW). In Table~\ref{tab:relwork-quantum}, the ``Obj.'' block gives the operator size $K\!\times\!N$ for power-aggregation objectives (as here and \cite{xiong2025das}) or ``---'' otherwise (sum-rate, SINR), with $\theta$ the phase alphabet; the design places no constraint on $K$ ($K\!=\!1$: SISO/MISO; $K\!>\!1$: $K$ users (MU) or one multi-antenna receiver). Algorithm tags abbreviate quantum-gradient descent (QGD); link tags use index modulation (IM), near-field (NF), full-duplex (FD). The ``Opt.''\ column marks how QAOA angles are obtained: \emph{online} schemes re-optimize per instance; \emph{train-once} schemes optimize once on a representative model and reuse them, but leave training cost and   scaling uncharacterized and exposed to barren plateaus~\cite{colella2025barren}; \emph{offline} (this work) fixes them by classically optimizing  deterministic infinite-size $V_p$ at  BGMZ cost $O(q_{\max}M_p\log M_p)$ per evaluation ($M_p=4^p$), with no variational training and only the final fixed-level circuit run on hardware. 
 
\subsubsection{Classical phase-only RIS optimizers}
Convex relaxations such as semi-definite relaxation (SDR) for non-convex constant-modulus optimization lift the constraint to semidefinite surrogates \cite{wuzhang2019sdr}. Discrete alphabets are typically handled by divide-and-sort alternating maximization (DaS-AM)~\cite{xiong2025das}. These methods, however, inherit  NP-hardness of the discrete unimodular quadratic problem~\cite{zhang2006complex}: relaxation followed by projection loses optimality while rounding, local searches certify only stationarity,  SISO guarantees do not extend to general $K\!\times\!N$ matrices, and the discrete space of $2^{B \,N}$ configurations rules out exhaustive verification beyond reduced sizes.
\subsubsection{Quantum-assisted RIS phase optimizers}
D-Wave annealing maps scattered power to a quadratic Ising model ~\cite{ross2022engineering}; extensions reach $177$ logical $1$-bit elements ($6400$ via hybrid decomposition)~\cite{lim2024taprls}, feedback-driven hybrid loops~\cite{ross2023nemo}, QUBO recipes~\cite{qaem4eng2025}, cell-free multi-bit QUBO~\cite{ohyama2022cellfree}, and $1$-bit index modulation~\cite{ris1bit2025}. Gate-based routes include $2$-body spin-glass QAOA at $N\!=\!9$, $p\!=\!18$~\cite{colella2024multipath} with   barren plateau analysis in \cite{colella2025barren},  physics-informed QUBOs~\cite{pasquale2026piqaoa}, and a QGNN on IBM $127$-qubit hardware~\cite{qgnn2026}; STAR-RIS schemes alternate QUBO phase steps with classical beamforming ($N\!\le\!18$)~\cite{pham2026starjoint}, tile QAOA over $8$-qubit blocks ($N\!=\!64$)~\cite{lim2026starcoupled}, add quantum natural gradients for NF STAR-RIS ($N\!=\!192$)~\cite{paul2026qaqgnf}, or use projected QGD \cite{sharma2026pqgd}. A quantum approximate-and-gradient hybrid optimizes NF holographic-MIMO apertures with joint phase and amplitude under a sum-rate objective~\cite{paul2026hmimo}, and quantum solvers address MIMO ML detection~\cite{gulbahar2025qaoa,norimoto2023hubo}. 
Closest in objective is the $b$-bit quantized MIMO beamforming of~\cite{mitsiou2025qaoabbit}, which maximizes $|\mathbf{A} \, \mathbf{f}|^{2}$ over a finite-alphabet phase vector but targets transmit/receive precoding and trains the angles online per instance (warm-started COBYLA), whereas ours are fixed offline. A direct binary \emph{phase} expansion $\theta_n\!=\!\sum_i x_{n,i}\theta_i$~\cite{mitsiou2025qaoabbit} makes the cost Hamiltonian $2 \,b$-body (their $1$-bit/QPSK case being the order-2 special instance). Our \emph{nonuniform} unit-circle dictionary instead represents each phasor by quadratic sign templates, an exact order-4 HUBO with an order-2 Ising surrogate, so the $L\!=\!2^{M}$ phase resolution is decoupled from the interaction order.  Order-2 design reduces gate-count, and the pure even-$q\!\ge\!4$ OGP obstruction~\cite{BassoFOCS22} does not cover the quadratic model.  Offline angles maximize the infinite-size depth-$p$ ensemble energy $V_p$ by construction (Section~\ref{sec:Vp}), so warm-started per-instance search recovers only a finite-size gap vanishing as $n\!\to\!\infty$; our channel-independent pack instead furnishes this initialization at no per-channel cost. In this work, we benchmark offline-angle construction against a classical solver for the discrete-phase objective (DaS-AM, Section~\ref{subsec:A-main-paper}) and exhaustive search at small $N$; an   online-versus-offline QAOA angle comparison is a separate question, beyond the present scope. Since offline angles use no structure of $A$, we validate the transfer on Rician/LOS, cascaded double-fading, and spatially-correlated RIS channels (Section~\ref{sec_simul}).

The remainder covers constant-level QAOA and $V_p$ (Sec.~\ref{sec:bg}), the hard-unit-circle dictionary and RIS phase formulation (Secs.~\ref{sec:binquad-angles}--\ref{sec:risphase}), the induced Hamiltonian and order-2 reduction (Sec.~\ref{sec:order2}), $V_p$ and local-field angle design (Sec.~\ref{sec:Vp}), channel models (Sec.~\ref{sec:ris-A-unified}), complexity (Sec.~\ref{sec:complexity}), simulations (Sec.~\ref{sec:sim}), and open issues and conclusions (Secs.~\ref{sec_open}--\ref{sec_con}).
\section{Background on  QAOA }
\label{sec:bg}
\label{secQAOA}
QAOA prepares a parameterized state by alternating the problem and mixing unitaries over $p$ layers~\cite{farhi2014quantum}, and asymptotic behavior is studied for random spin models when $p$ remains fixed~\cite{farhi2022quantum,BassoFOCS22}.  It provides a theoretical benchmark that can be compared against the discrete RIS models developed later through the $V_p$ formalism of mixed-$q$ Hamiltonians. A combinatorial problem is written as an Ising cost as follows:
\begin{equation}
\label{eq:costfunc}
C(\boldsymbol z)=\sum_{i<j} J_{ij}\,z_i\,z_j+\sum_{j} h_j\,z_j ,
\end{equation}
over $n$-bit spin strings $\boldsymbol z=(z_1,\dots,z_n)\in\{+1,-1\}^n$, with quadratic couplings $J_{ij}$ and local fields $h_j$. The corresponding cost Hamiltonian is $H_C=\sum_{i<j} J_{ij}\,Z_i Z_j+\sum_{j} h_j\,Z_j$, with $Z_j$ the Pauli-$Z$ operator on qubit $j$. QAOA prepares the parameterized state $\ket{\Psi(\boldsymbol\gamma,\boldsymbol\beta)}=\big(\prod_{\ell=1}^{p} U_B(\beta_\ell)\,U_C(\gamma_\ell)\big)\,\ket{+}^{\otimes n}$ from  uniform superposition $\ket{+}^{\otimes n}$ 
where alternating  cost  $U_C(\gamma)=e^{-i\gamma H_C}$ and  mixing  $U_B(\beta)=e^{-i\beta B}$ unitaries with mixer $B=\sum_{j} X_j$ ($X_j$ the Pauli-$X$ operator) over $p$ layers. The $2 \,p$ angles $\boldsymbol\Theta\equiv(\boldsymbol\gamma,\boldsymbol\beta)$ are optimized so that $\braket{C}\equiv\braket{\Psi(\boldsymbol\gamma,\boldsymbol\beta)|H_C|\Psi(\boldsymbol\gamma,\boldsymbol\beta)}$ reaches its optimum and measured bitstrings concentrate on near-optimal $C(\boldsymbol z)$.  
\section{Binary--Quadratic Discrete Angle Synthesis}
\label{sec:binquad-angles}
\begin{algorithm}[!t]
\caption{Hard-Unit-Circle Angle Dictionary (Multi-Start)}
\label{alg:hardcircle}
\begin{algorithmic}[1]
\State \textbf{Input:} $\mathcal{D}\!\in\!\{\!-1,1\!\}^{L\times M}$; weights $\mu_1,\mu_2,\mu_3$; barrier $\varepsilon\!>\!0$; runs $N_o$; starts $N_{ms}$.
\State $(\delta^\star,s^\star)\leftarrow(\infty,\infty)$.
\For{$r=1,\dots,N_o$}
  \State Draw random small $(\mathbf{l}_c,\mathbf{l}_s,\mathbf{Q}_c,\mathbf{Q}_s)$
  \State Min. \eqref{eq:angle-obj} from $N_{ms}$ starts; keep best feasible.
  \State $\delta\leftarrow\max_i|x_i^2{+}y_i^2{-}1|$,\quad $s\leftarrow\bigl(\tfrac{1}{L}\textstyle\sum_i(\Delta\theta_i{-}\tfrac{2\pi}{L})^2\bigr)^{1/2}$.
  \If{$\delta<\delta^\star$, or $\delta{=}\delta^\star$ and $s<s^\star$}
    \State save $(\mathbf{l}_c,\mathbf{l}_s,\mathbf{Q}_c,\mathbf{Q}_s)$ as best; $(\delta^\star,s^\star)\leftarrow(\delta,s)$.
  \EndIf
\EndFor
\State \textbf{Output:} best $(\mathbf{l}_c,\mathbf{l}_s,\mathbf{Q}_c,\mathbf{Q}_s)$ and $\theta_i=\operatorname{atan2}(y_i,x_i)$.
\end{algorithmic}
\end{algorithm}
In Algorithm~\ref{alg:hardcircle}, we present the hard-unit-circle angle dictionary used throughout: a multi-start synthesis that fits linear and quadratic sign-pattern templates to points on the unit circle with near-uniform angular spacing. Assume that $M\!\ge\!1$ and $L\!=\!2^M$. Stack all $\{\!-1,1\!\}$ sign patterns as rows of
$\mathcal{D}\!\in\!\{\!-1,1\!\}^{L\times M}$; denote the $i$th row by $c_i^\top$.
We synthesize
$
(\cos\theta_i,\sin\theta_i) \equiv (x_i,y_i)$ and $
\theta_i=\operatorname{atan2}(y_i,x_i),
$
using \emph{linear} templates $\mathbf{l}_c,\mathbf{l}_s\in\mathbb{R}^M$ and \emph{quadratic} $\mathbf{Q}_c,\mathbf{Q}_s\in\mathbb{R}^{M\times M}$:
\begin{equation}
x_i = c_i^\top \mathbf{l}_c + c_i^\top \mathbf{Q}_c\,c_i,\qquad
y_i = c_i^\top \mathbf{l}_s + c_i^\top \mathbf{Q}_s\,c_i.
\label{eq:model-xy}
\end{equation}
where  $\mathbf{Q}_c=\mathbf{Q}_c^\top$, $\mathbf{Q}_s=\mathbf{Q}_s^\top$ and
$\mathrm{diag}(\mathbf{Q}_c)=\mathrm{diag}(\mathbf{Q}_s)=\mathbf{0}$. We enforce the exact unit-norm constraint as follows:
\begin{equation}
x_i^2 + y_i^2 \;=\; 1,\qquad i=1,\dots,L,
\label{eq:unit-eq}
\end{equation}
Let $\theta_i$ be wrapped to $[0,2\pi)$ and  sorted angles be
$\{\theta_{(i)}\}_{i=1}^L$ with wrap-around gaps
$
\Delta\theta_i := \theta_{(i+1)}-\theta_{(i)}$ and
$\theta_{(L+1)} := \theta_{(1)}+2\pi$.
We target equal spacing $2\pi/L$ with a quadratic gap loss and a smooth log-repulsion against coincident angles:
\begin{equation}
\min_{\mathbf{l}_c,\mathbf{l}_s,\mathbf{Q}_c,\mathbf{Q}_s}
 \mu_1\sum_{i=1}^L\!\bigl(\Delta\theta_i - \tfrac{2\pi}{L}\bigr)^2
- \, \mu_2\sum_{i=1}^L\!\log(\Delta\theta_i+\varepsilon)
 + \mu_3\,\mathcal{R}
\label{eq:angle-obj}
\end{equation}
subject to \eqref{eq:model-xy},\eqref{eq:unit-eq} where $\mu_1,\mu_2,\mu_3\!\ge\!0$ are weights, $\varepsilon\!>\!0$ protects the barrier at zero,
and $\mathcal{R}\equiv\|\mathbf{l}_c\|_2^2{+}\|\mathbf{l}_s\|_2^2
{+}\|\mathbf{Q}_c\|_F^2{+}\|\mathbf{Q}_s\|_F^2$ keeps
template coefficients bounded.
The log term spreads points uniformly on the circle; we solve \eqref{eq:angle-obj} by multi-start  sequential quadratic programming. Each objective evaluation is $O(L \,M^2)$ due to the  forms in \eqref{eq:model-xy}.
\section{RIS Phase Maximization}
\label{sec:risphase}
Let $T\in\{-1,1\}^{N\times M}$ be the optimization variable, its $j$th row $c_j^\top$ encoding element $j\in\{1,\dots,N\}$. Given $\mathbf{l}_c,\mathbf{l}_s,\mathbf{Q}_c,\mathbf{Q}_s$ from Algorithm~\ref{alg:hardcircle}, set $x_j(T)=c_j^\top \mathbf{l}_c + c_j^\top \mathbf{Q}_c\,c_j$ and $y_j(T)=c_j^\top \mathbf{l}_s + c_j^\top \mathbf{Q}_s\,c_j$ as in \eqref{eq:model-xy}, with $\theta_j(T)=\operatorname{atan2}(y_j,x_j)$ and unit modulus $x_j(T)^2+y_j(T)^2=1$, so that
\begin{equation}
z_j(T) = e^{i\theta_j(T)} \;=\; x_j(T) + i\,y_j(T).
\label{eq:thetazC}
\end{equation}
Stacking into $z(T)=x(T)+i\,y(T)$ over the feasible set $\Delta=\{\theta(T)\in[0,2\pi)^N: T\in\{-1,1\}^{N\times M},\ x_j^2+y_j^2=1,\ j=1,\dots,N\}$, and with $A=A_R+iA_I\in\mathbb{C}^{K\times N}$ ($A_R,A_I\in\mathbb{R}^{K\times N}$), the power-maximization problem is the following:
\begin{equation}
\max_{\theta\in\Delta}\|A\,e^{i\theta}\|_2^2=\max_{T\in\{-1,1\}^{N\times M}}\|A\,z(T)\|_2^2 ,
\end{equation}
which expands as follows:
\begin{equation}
\label{eq:blockquad}
\|A\,z(T)\|_2^2 = x^\top G\,x + y^\top G\,y + 2\,x^\top S\,y ,
\end{equation}
with $G \equiv A_R^\top A_R + A_I^\top A_I$ and $S \equiv A_I^\top A_R - A_R^\top A_I$, a discrete binary-quadratic program over $T$. 
\section{Spin Hamiltonian and the Order-2 Reduction}
\label{sec:order2}

\begin{figure*}[!t]
\centering
\includegraphics[width=7in]{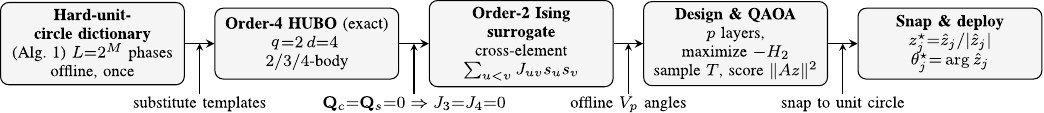}
\caption{End-to-end system design. A single offline hard-unit-circle dictionary (Algorithm~\ref{alg:hardcircle}), $L{=}2^{M}$ phases, induces the exact order-4 HUBO~\eqref{eq:ising4}; switching off the quadratic templates ($\mathbf{Q}_c{=}\mathbf{Q}_s{=}\mathbf{0}$) zeros the cubic and quartic couplings, leaving the order-2 Ising surrogate~\eqref{eq:q2H}, which a channel-independent $V_p$-designed QAOA angle pack (Section~\ref{sec:Vp}) optimizes; the sampled sign patterns are snapped to the unit circle for deployment.}
\label{fig:pipeline}
\end{figure*}

The end-to-end system design is summarized in Fig.~\ref{fig:pipeline}: a single offline hard-unit-circle dictionary induces an exact order-4 HUBO, whose order-2 reduction is optimized by a channel-independent QAOA angle pack and snapped back to the unit circle. Section~\ref{sec:risphase} defines the induced cost
$\mathcal H(T)=-\|Az(T)\|_2^2$. Write
$s_{j,i}\equiv[T]_{j,i}\in\{-1,1\}$ for the $i$th entry of row $c_j^\top$ of $T$, and flatten the pair into a single spin label
$u\equiv(j,i)$: the $n=N\,M$ entries of $T$ are the spins $s_u=s_{j,i}$. Expanding
$\mathcal H(T)$ in these variables and reducing powers with $s_{j,i}^2=1$ gives the following
mixed-order HUBO model:
\begin{equation}
\label{eq:ising4}
\begin{aligned}
\mathcal H(T)=-C_0&+\sum_{u<v}J_{uv}s_us_v+\sum_{u<v<w}J_{uvw}s_us_vs_w\\
&+\sum_{u<v<w<x}J_{uvwx}s_us_vs_ws_x ,
\end{aligned}
\end{equation}
Since $x_j^2+y_j^2=1$ for every Boolean row, all same-element contributions collapse into the constant $-\operatorname{tr}(G)$, leaving only cross-element couplings. $C_0\equiv\operatorname{tr}(G)=\|A\|_F^2$ is the state-independent offset of the power cost $C(z)\equiv\|Az(T)\|_2^2=-\mathcal H(T)$. It is the incoherent random-phase power floor, and $C_{\mathrm{opt}}-C_0$ is the coherent excess available to the optimizer. Since $x_j,y_j$ are degree-$d$ in the sign pattern $\mathbf{c}_j$ and the power
is bilinear in $(x,y)$, the induced Hamiltonian has order $q=2 \,d$. The quadratic templates ($d=2$) generate the cubic
and quartic  terms (explicit coefficients in the reproducibility package~\cite{repository}). Therefore,
setting
$\mathbf{Q}_c=\mathbf{Q}_s=\mathbf{0}$ leads to $3$ and $4$-body terms $
J_{\text{3}}=J_{\text{4}}=0$
 leaving the order-2 Hamiltonian built from the offset $C_0$ and the couplings \eqref{eq:J2} as follows:
\begin{equation}
H_{2}(T)=-\operatorname{tr}(G)+\sum_{u<v}J_{uv}\,s_u s_v ,
\label{eq:q2H}
\end{equation}
where cross-element coupling is the following for $j\ne k$:
\begin{equation}
\label{eq:J2}
\begin{aligned}
J_{(j,m)(k,n)}={}&-2G_{jk}\!\left(\ell_{c,m}\ell_{c,n}+\ell_{s,m}\ell_{s,n}\right)\\
&-2S_{jk}\ell_{c,m}\ell_{s,n}-2S_{kj}\ell_{s,m}\ell_{c,n}
\end{aligned}
\end{equation}
Linear images fall slightly off the unit circle ($x_j^2{+}y_j^2\neq1$) with the quadratic templates off, so \eqref{eq:q2H} is a \emph{surrogate}: it keeps the offset $-\operatorname{tr}(G)$ and the cross-element couplings $J_{uv}$ but drops the same-element terms an exact off-circle expansion would carry. The design-and-snap projection removes this deviation.

\subsection{Unit-circle realizability of the linear dictionary}
Linear synthesis gives
$x_j=\mathbf{c}_j^{\!\top}\boldsymbol{\ell}_c$ and
$y_j=\mathbf{c}_j^{\!\top}\boldsymbol{\ell}_s$. The unit-circle equalities
\eqref{eq:unit-eq} can still be \emph{imposed} on the template optimization
(with the quadratic templates active they are met for all $2^M$ patterns to
$\sim\!10^{-13}$, Section~\ref{subsec:framework}), but the linear model cannot
\emph{satisfy} them beyond a degenerate solution:
$x_j^2+y_j^2=\mathbf{c}_j^{\!\top}(\boldsymbol{\ell}_c\boldsymbol{\ell}_c^{\!\top}
+\boldsymbol{\ell}_s\boldsymbol{\ell}_s^{\!\top})\mathbf{c}_j$ is a rank-two
quadratic form, and demanding the value $1$ on \emph{every} sign pattern forces
the Gram matrix diagonal, leaving at most two active sign bits. Hence the linear
dictionary realizes \emph{at most four} on-circle phases regardless of $M$ (exactly four in the nondegenerate rank-two case): it is
a $2$-bit RIS. Numerically, requiring $|x_j^2+y_j^2-1|\le1\%$ admits only four
phases even at $M=5,6$, and a $5\%$ tolerance admits on the order of $8$--$20$. The image points occur in antipodal pairs
($c\to-c$ sends $(x,y)\to(-x,-y)$), reflecting global-phase
symmetry.
The linear model is thus a $2$-bit RIS plus a controlled
off-circle relaxation.
\subsection{Order-2 reformulation by design-and-snap}
\begin{figure}[!t]
\centering
\includegraphics[width=3.25in]{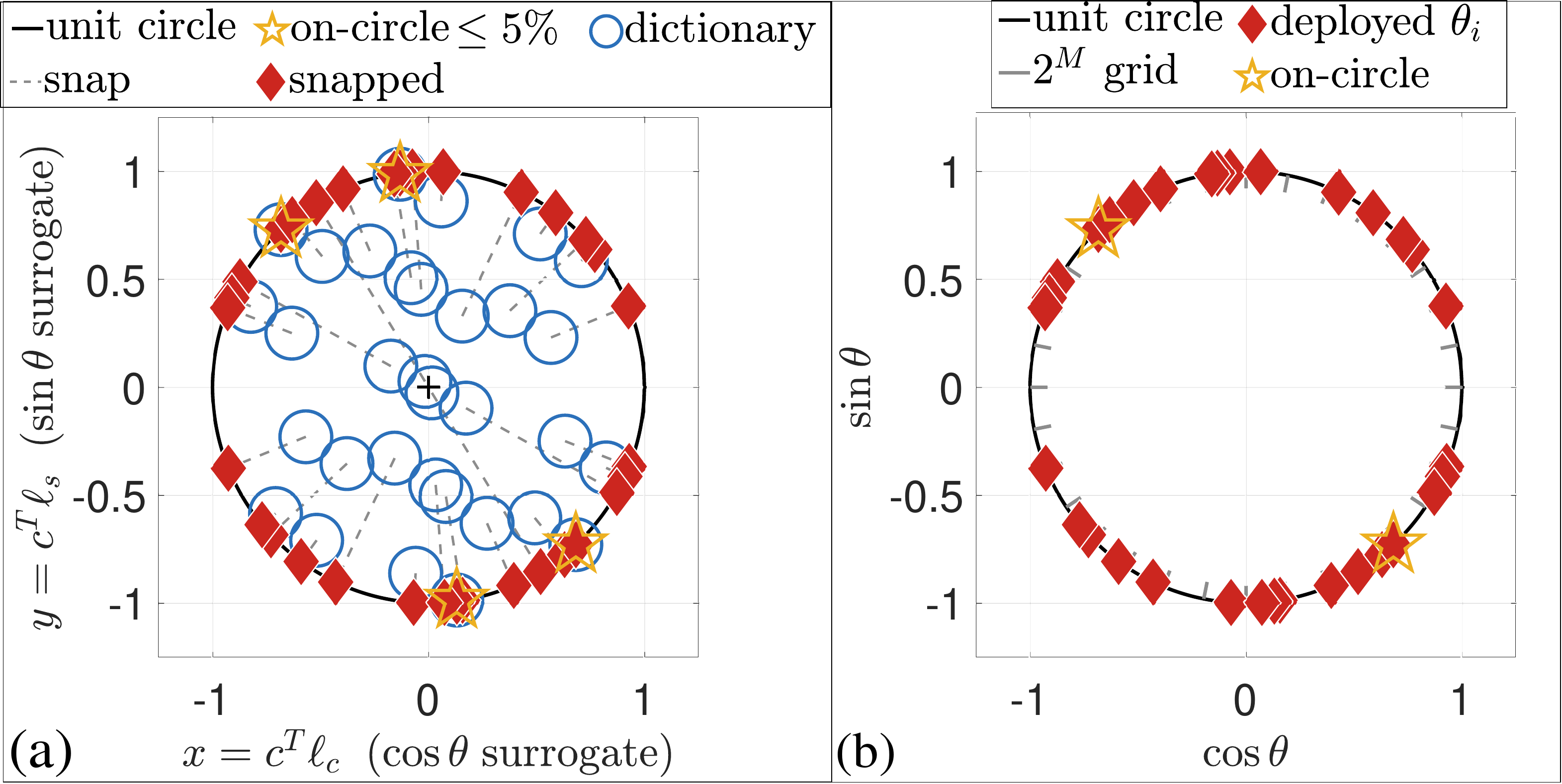}
\caption{Order-2 dictionary, $M{=}5$. (a)~The $32$ linear images (open) lie off
the unit circle; the snap deploys $\theta$ (filled); stars: on-circle to
$5\%/1\%$. (b)~Deployed $32$-phase alphabet vs.\ uniform $5$-bit grid (ticks).}
\label{fig:phase-disc}
\end{figure}

We deploy a two-step
\emph{design-and-snap} procedure as shown in Fig.~\ref{fig:phase-disc}: (i)~\emph{design}, maximize the order-2 surrogate
$-H_2(T)$ over $T\in\{\pm1\}^{N\times M}$, giving off-circle pairs
$\hat z_j=\hat x_j+i\hat y_j$; (ii)~\emph{snap}, project each element to the unit circle,
$z_j^{\star}=\hat z_j/|\hat z_j|$, i.e., deploy $\theta_j^{\star}=\arg\hat z_j$.
Algorithm~\ref{alg:hardcircle} is not re-run for this model: the order-2 synthesis
reuses the linear templates $\boldsymbol{\ell}_c,\boldsymbol{\ell}_s$ of the single
offline dictionary with $\mathbf{Q}_c,\mathbf{Q}_s$ switched off; the runtime optimization over
$T$ is unconstrained. The projection preserves the synthesized angle exactly; the only approximation is that the
design step scored configurations with the off-circle surrogate rather than the true
on-circle power, and the two generally differ by a term linear in the radial deviations plus higher-order corrections, a mismatch that is empirically small for the learned dictionary. Empirically, on the ten $N=16$ channels of Section~\ref{sec:sim} the design-and-snap phases lose about $
0.1$ dB versus the continuous-phase optimum.
Against the order-2 reference $C_{\mathrm{opt}}$ (the multi-start single-flip local search of  Section~\ref{subsec:A-main-paper}; the certified optimum where $N\!\le\!6$), we report the best-sample ratio $r_b\equiv C_{\mathrm{best}}/C_{\mathrm{opt}}$ and the mean ratio $r_m\equiv\langle C\rangle/C_{\mathrm{opt}}$. Two costs enter: the surrogate Ising $-H_2(T)$ that the circuit optimizes and the deployed snapped power $\|Az^\star(T)\|_2^2$.   $r_m$ and the $V_p$ benchmark (Section~\ref{sec:vp_benchmark}) are evaluated on $-H_2$, the quantity $V_p$ predicts, whereas  $r_b$ is evaluated on the deployed power; in each case $C_{\mathrm{opt}}$ is scored on the same cost as its numerator (the reference configuration is snapped before physical scoring), the two denominators differing only by the $\approx0.1$ dB snap loss. At $N\!\le\!6$ exhaustive enumeration is reported on both costs, confirming the surrogate and physical optima coincide within this gap.
\section{The Infinite-Size Energy $V_p$ and Angle Design}
\label{sec:Vp}
The angles in this paper are designed against $V_p$, the
infinite-size energy of constant-level QAOA on a Gaussian mixed-$q$ ensemble
$\mathcal G$, computed by the fixed-point iteration of
BGMZ \cite{BassoFOCS22} (see also
\cite{farhi2022quantum,Basso2021}). The iteration is restated in Appendix~\ref{app:vp-iter}, where we also derive its fast Walsh--Hadamard transform (FWHT) evaluation, recasting the multi-fold sums as XOR-convolution powers (introduced here) to reduce each sweep to $O(q_{\max}M_p\log M_p)$, $M_p=4^p$; our implementation evaluates it verbatim. The design needs only that $V_p$ is a deterministic function of $(\gamma,\beta)$ and  $c=(c_1,\dots,c_{q_{\max}})$. The RIS matrix $A$ is generated separately in Section~\ref{sec:ris-A-unified}.

\subsection{Local-field matching for angle transfer}
\label{subsec:local-field-matched-cq}
The angle pack is optimized once on the Gaussian surrogate of Appendix~\ref{app:vp-iter} and reused across instances; for this reuse to be accurate at finite depth, the surrogate must reproduce the typical local field on one spin. We  formalize this by writing   \eqref{eq:ising4} as
$\mathcal H(T)=-C_0+\sum_{q=2}^{4}\sum_{e\in\mathcal I_q}J^{(q)}_{e}\prod_{u\in e}s_u$ where $\mathcal I_q$ is the set of order-$q$ \emph{hyperedges} (index tuples 
$u < v$, $u < v < w $, and $u < v < w< x$ carrying a nonzero coupling) and $J^{(q)}_{e}$ is
the coefficient of hyperedge $e$, i.e., $J_{uv}$, $J_{uvw}$, $J_{uvwx}$ of
\eqref{eq:ising4}. Let $n =N \, M$. For $q\in\{2,3,4\}$, define
\begin{equation}
\overline V_q= (1 \, / \, |\mathcal I_q|) \sum_{e\in\mathcal I_q}\bigl(J^{(q)}_{e}\bigr)^{2},\qquad
D_q=q\,|\mathcal I_q| \, / \, n 
\label{eq:VqDq-local-v15}
\end{equation}
$\overline V_q$ is the mean-square coupling and $D_q$ the average number of order-$q$ hyperedges per spin ($q\,|\mathcal I_q|$ incidences over $n$ spins). Their product is the per-spin local-field variance: the order-$q$ field $h^{(q)}_u=\sum_{e\in\mathcal I_q:\,u\in e}J^{(q)}_{e}\prod_{v\in e\setminus\{u\}}s_v$ collects the terms linear in $s_u$ in \eqref{eq:ising4}, and under uniformly random signs $\langle\mathrm{Var}\,h^{(q)}_u\rangle=D_q\overline V_q$, the quantity the surrogate must reproduce.
The surrogate coefficients are fixed by carrying this scale into the ensemble
convention of Appendix~\ref{app:vp-iter}: $c_1=0$ and
\begin{equation}
 c_q=\sqrt{(q-1)!\,D_q\overline V_q},\qquad q\in\{2,3,4\},
\label{eq:cq-local-factorial-v15}
\end{equation} 
For the order-2 models, the rule sets $c_2^{2}=D_2\overline V_2=\langle\mathrm{Var}\,h^{(2)}_u\rangle$, the per-spin local-field variance in the \emph{unordered-pair} normalization adopted here; this convention is validated directly by the exact $p{=}1$ and PPS $p{=}2$ checks of Table~\ref{tab:vp_valid}. The factor $(q-1)!$ at $q=3,4$ is multiplicity bookkeeping between the unordered hyperedge sum of \eqref{eq:ising4} and the ordered-tuple Gaussian convention of Appendix~\ref{app:vp-iter}, where one unordered order-$q$ edge corresponds to $q!$ ordered tuples. This factorial extension to $q{=}3,4$ is a provisional moment-matching surrogate rather than a derived identity: the RIS couplings are permutation-tied and correlated through the common $G,S$, unlike the independent ordered tensors of the reference Gaussian ensemble. Only the $q{=}2$ calibration is validated here (Table~\ref{tab:vp_valid}).
 
\subsection{Angle optimization and validation workflow}
\label{subsec:angle-optimization-local-v15}
We optimize $V_p$ over $\boldsymbol\gamma,\boldsymbol\beta$ by multi-start local optimization, accepting the largest $V_p$ (the start count is an optimization parameter). The phase separator acts on $C_{\mathrm{power}}\equiv\|Az\|_2^2=-\mathcal H$ (the circuit \emph{maximizes} power, matching the BGMZ convention): 
$U_C(\gamma)=e^{-i\gamma C_{\mathrm{power}}}$, $U_B(\beta)=e^{-i\beta\sum_i X_i}$, so the $V_p$ angles transfer unchanged at the design scale $c_{2,\mathrm{ref}}{=}0.300$; an instance whose local-field scale $c_2$ differs re-matches it by rescaling only the cost angle, $\gamma_{\rm circuit}=\gamma_{V_p}\,c_{2,\mathrm{ref}}/c_2$ (the mixer $\beta$ is scale-free), as used  in Section~\ref{subsec:phys-robust}.
\section{RIS Channel Matrix for the $\ell_2$ Objective}
\label{sec:ris-A-unified}
We study the phase-only power aggregation problem:
\begin{equation}
\label{eq:obj-l2}
\max_{\Omega\in[0,2\pi)^N}\ \big\|A\,u(\Omega)\big\|_2^2,
\qquad
u(\Omega)=\exp\!\big(\mathrm{j}\,\Omega\big)\in\mathbb{C}^N,
\end{equation}
where $A\!\in\!\mathbb{C}^{K\times N}$ maps RIS element phases to $K$ observation ports. The choice of $A$ governs realism and how well \eqref{eq:obj-l2} proxies received power \cite{ElMossallamy2020}.  
Consider a narrowband downlink in which a base station (BS) with $N_t$
antennas serves $K$ receive (observation) ports through an $N$-element RIS,
the direct BS-receiver path being blocked or absorbed into the noise. Let
$H_{\!RB}\in\mathbb{C}^{N\times N_t}$ and
$H_{\!UR}\in\mathbb{C}^{K\times N}$ denote the BS--RIS and RIS-receiver
channel matrices, $W\in\mathbb{C}^{N_t}$ the transmit precoder carrying the
unit-power data symbol $s\in\mathbb{C}$  and $n\in\mathbb{C}^{K}$ additive noise. The received
signal and the induced channel matrix are as follows:
\begin{equation}
\label{eq:A-cascade}
y = H_{\!UR}\,\mathrm{diag}(u)\,H_{\!RB}\,W\,s + n; \,\, A = H_{\!UR}\,\mathrm{diag}(z) 
\end{equation}
where $z = H_{\!RB}\,W\,s\in\mathbb{C}^N$ is the effective per-element
incident excitation. Since $\mathrm{diag}(u)\,z=\mathrm{diag}(z)\,u$, the
received signal is $y=Au+n$, so \eqref{eq:obj-l2} is exactly the received
power.  In stochastic form the baseline is rich-scattering or LOS:
\begin{equation}
\label{eq:A-stat}
A_{mn}\sim\mathcal{CN}(0,\sigma^2)\;\;\text{(i.i.d.)}\;\;\text{or}\;\;
A=\sqrt{\tfrac{\kappa}{\kappa+1}}\,\bar{A}+\sqrt{\tfrac{1}{\kappa+1}}\,A_{\mathrm{iid}},
\end{equation}
where $A_{\mathrm{iid}}$ has i.i.d.\ $\mathcal{CN}(0,\sigma^2)$ entries as in the first branch, $\bar{A}$ a deterministic LOS component, and $\kappa\ge 0$ the Rician factor; by default the campaigns of Section~\ref{sec:sim} use the i.i.d.\ branch with $\sigma=1/\sqrt{N}$ and $K=N$ (Section~\ref{subsec:framework}). Since the dictionary synthesis and HUBO reduction of Sections~\ref{sec:risphase}-\ref{sec:Vp} use no structure of $A$, any fixed $A\in\mathbb{C}^{K\times N}$, e.g., far-field steering manifolds, near-field XL-RIS generalizations, or multiport mutual-coupling models~\cite{ElMossallamy2020,Nerini2024}, drops into \eqref{eq:obj-l2} unchanged.
To probe robustness beyond i.i.d.\ Rayleigh, we exercise four ensembles spanning the rich-scattering, LOS, cascaded, and spatially-correlated regimes, all Frobenius-normalized to $\mathbb{E}\|A\|_F^2=K$ so that only channel \emph{structure}, not energy, varies across families. \emph{(M1) i.i.d.\ Rayleigh}: the structure-free baseline $A_{mn}\sim\mathcal{CN}(0,1/N)$ of \eqref{eq:A-stat}, zero spatial correlation, no specular component. \emph{(M2) Rician/LOS}: the second branch of \eqref{eq:A-stat} with a rank-one specular term $\bar A=\mathbf a_K(\vartheta_r)\,\mathbf a_N(\vartheta_t)^{\mathsf H}$ over the diffuse  component~\cite{Zhang2023StatCSI}, $\kappa\in\{3,10\}$~dB  at receive/transmit angles $\vartheta_r,\vartheta_t$ and uniform-linear-array steering vectors $[\mathbf a(\vartheta)]_k=e^{\mathrm{j}\pi k\sin\vartheta}$ with $\|\bar A\|_F^2=K$. \emph{(M3) Cascaded double-fading}: the two-hop reduction \eqref{eq:A-cascade}, $A=H_{\!UR}\,\mathrm{diag}(z)$ with $z=H_{\!RB}\,W$, i.i.d.\ Rayleigh hops, and a random unit-norm precoder $W$ \cite{wuzhang2019sdr}, so every column shares the common per-element excitation $z_n$; each entry is a product of two complex Gaussians, jointly normalized by the sample RMS of $z$ (hence weakly dependent) and heavier-tailed than M1. \emph{(M4) Spatially-correlated}: at sub-wavelength spacing the i.i.d.\ model is non-physical for a densely-packed array under isotropic scattering~\cite{bjornson2021rayleigh}, with the following correlation:
\begin{equation}
\label{eq:A-corr}
[R]_{nm}=\operatorname{sinc}\!\big(2\|u_n-u_m\|/\lambda\big),\qquad
\operatorname{sinc}(x)=\frac{\sin\pi x}{\pi x}
\end{equation}
over the array ($u_n$ the element positions; a uniform linear array at spacing $d=\lambda/4$), giving $A=A_{\mathrm{iid}}\,R^{1/2}$ (or the same correlation on both hops of \eqref{eq:A-cascade}), with $R$ low-rank. 
\subsection{Discrete-Phase Power Aggregation (DaS-AM)}
\label{subsec:A-main-paper}
The DaS-AM framework of \cite{xiong2025das} maximizes the $\ell_2$
power-aggregation objective with discrete per-element phases,
\begin{equation}
\max_{\Omega\in\Delta^N}\ \big\|A\,e^{\mathrm{j}\Omega}\big\|_2, \,\,
\Delta=\{0,\delta,\ldots,(2^B\!-\!1)\delta\},\ \delta=\tfrac{2\pi}{2^B},
\label{eq:Q2-main}
\end{equation}
and shows that a broad class of RIS-aided links reduce to \eqref{eq:Q2-main} via an explicit $A$. Summing received powers over $K$ users ($\Phi_k$) gives a Hermitian form $\sum_k\Phi_k^{\!H}\Phi_k\succeq0$; any $A$ with $A^{\!H}A=\sum_k\Phi_k^{\!H}\Phi_k$ recovers \eqref{eq:Q2-main}, so the $K\times N$ operator already covers the multi-user case. For the order-2 surrogate of Section~\ref{sec:order2} we use a $B{=}1$ solver in the spirit of this inner step: a multi-start single-flip best-improvement local search with random, spectral, and unimodular-lifting  initializations, each polished to a single-flip fixed point. This solver is not a line-by-line implementation of \cite{xiong2025das}, and it does not certify global optimality, since single-block exactness and monotone alternation do not certify the joint nonconvex $\ell_2$ optimum. It reproduces the exhaustive optimum at $N\!=\!5,6$ and improves on a single start by under $0.27\%$ on the first $N\!=\!24$ channel, so we use its value as the certified optimum $C_{\mathrm{opt}}$ at $N\!\le\!6$ and as an order-2 reference (Section~\ref{sec:q2qaoa}).
\section{Computational Complexity}
\label{sec:complexity}
Table~\ref{tab:complexity} collects the operation counts (derivations in Appendix~\ref{app:complexity}), separating   one-time \emph{offline} design, dictionary and $V_p$ angles, both independent of $N$, from the \emph{per-instance} cost. $N_{\mathrm{eval}}$ and $N_{\mathrm{sweep}}$ are the numbers of multi-start evaluations of  dictionary objective~\eqref{eq:angle-obj} and of $V_p$ fixed-point sweeps run to convergence, $N_s$ the number of QAOA shots, and $N_r$ and $N_{\rm flip}$ the number of restarts and accepted single-flips of the classical local-search reference; one evaluation costs $O(LM^2)$ and one sweep $O(q_{\max}p\,4^p)$, both set by the optimizer rather than by $N$;  full $V_p$ pack further multiplies the per-sweep $O(q_{\max}p\,4^p)$ by the sweep count and the $(\gamma,\beta)$ multi-start size, all $N$-independent, so the offline totals stay $N$-independent. At deployment a surface pays only the per-instance HUBO assembly, the dense order-2 circuit, and the classical scoring of the $N_s$ samples on $\|Az\|_2^2$, comparable to polynomial-time classical solvers~\cite{xiong2025das} and  with no per-channel variational training. The gate count assumes all-to-all connectivity; the \textsc{swap} overhead   is left to the hardware study of Section~\ref{sec_open}. The exact finite-size $p{=}1$ RIS correlator used as an anchor needs one $O(n^{3})$ evaluation per size; the instance-free BGMZ $V_p$ itself is the $O(q_{\max}p\,4^p)$ fixed point above.
\begin{table}[t]
\centering\footnotesize
\caption{Offline-design and per-instance operation counts}
\label{tab:complexity}
\setlength{\tabcolsep}{3pt}
\renewcommand{\arraystretch}{1.15}
\begin{tabular}{|l|l|c|c|}
\hline
Phase & Operation & Classical, $O(\cdot)$ & Quantum, $O(\cdot)$\\
\hline\hline
\multirow{2}{*}{Offline} & Dictionary (Alg.~\ref{alg:hardcircle}) & $N_{\mathrm{eval}}\,LM^{2}$ & ---\\
\cline{2-4}
 & $V_p$ pack, per $p$ & $N_{\mathrm{sweep}}\,q_{\max}p\,4^{p}$ & ---\\
\multirow{2}{*}{Per inst.} & HUBO ($G,S,J$) & $KN^{2}\!+\!n^{2}$ & ---\\
\cline{2-4}
 & QAOA sample$+$score & $N_s KN$ & $N_s n^{2}p$\\
\hline
\multirow{2}{*}{Ref.} & $1$-flip reference & $N_r \,(n^{2}{+} \, n \,N_{\rm flip})$ & ---\\
\cline{2-4}
 & Exact $p{=}1$ RIS anchor & $n^{3}$ & ---\\
\hline
\end{tabular}
\end{table}
\section{Simulations}
\label{sec:sim}
\label{sec_simul}
We validate the offline angles in two tiers. We mostly characterize  the approach on the i.i.d.\ Rayleigh ensemble, the rich-scattering baseline of \eqref{eq:A-stat}, across the order-4 and order-2 models, depths $p=1$-$9$, and sizes $N=5$-$100$, with exact SV, MPS, PPS and closed-form methods cross-checking one another. The channel-robustness study of Section~\ref{subsec:phys-robust} then transfers the same channel-independent angles to Rician/LOS, cascaded double-fading and spatially-correlated RIS channels confirming the offline design is not tied to the i.i.d.\ assumption. Table~\ref{tab:sim-overview} lists the studies and their parameters.
\begin{table}[!t]
\centering\footnotesize
\caption{Overview of the simulation studies ($n = 5\,N$)}
\label{tab:sim-overview}
\setlength{\tabcolsep}{3pt}
\renewcommand{\arraystretch}{1.0}
\begin{tabular}{|c|l|l|c|l|c|}
\hline
\textbf{Order} & \textbf{Metric} & \textbf{$N$} & \textbf{$p$} & \textbf{Backend} & \textbf{\# Inst.}\\
\hline\hline
4 & $r_b$ & $12$ & $1$--$5$ & MPS & $10$\\\hline
\multirow{7}{*}{2} & $V_p$ pack & --- & $1$--$9$ & fixed-point & ---\\\cline{2-6}
 & $r_b,\ r_m$ & $16$ & $1$--$7$ & MPS & $10$\\\cline{2-6}
  & $r_m,\ r_m^{n}$ & $16,24;\,10$--$100$ & $1$--$9$ & $V_p$/PPS/exact & $10$\\\cline{2-6}
 & $r_m$ & $5$--$12$ & $1$--$3$ & SV/MPS & $100/50$\\\cline{2-6}
 & $P_\alpha$ & $5$--$12$ & $1$--$3$ & SV/MPS & $100/50$\\\cline{2-6}
 & $r_b,r_m$ (M2--M4) & $12$ & $1$--$4$ & MPS & $10$\\\cline{2-6}
& $r_b,r_m$ (M2--M4) & $5$ & $1$--$7$ & SV & $10$\\\hline
\end{tabular}
\end{table}
\subsection{Framework and protocol}
\label{subsec:framework}
\subsubsection{Channel instances}
The baseline campaigns draw $A\in\mathbb{C}^{K\times N}$ from one i.i.d.\ Rayleigh family (the i.i.d.\ branch of \eqref{eq:A-stat}, $\sigma=1/\sqrt{N}$): $A=(G_1+\mathrm{j}G_2)/\sqrt{2N}$, standard Gaussian $G_1,G_2$, $K=N$, with a common seed family so the small-$N$ checks, the $N=16$/$24$ campaigns, and the $N=5$--$100$ sweeps share one ensemble. Then $\mathbb{E}\|A\|_F^2=K$ and $\|A\|_F^2$ concentrates tightly, so power values are comparable across $N$. The Rician/LoS, cascaded~\eqref{eq:A-cascade}, and spatially-correlated constructions of Section~\ref{sec:ris-A-unified} are exercised in the channel-robustness study of Section~\ref{subsec:phys-robust}.
\subsubsection{Angle Dictionary}
The order-2 dictionary of Section~\ref{sec:binquad-angles} (Fig.~\ref{fig:phase-disc}(b)) is synthesized offline by Algorithm~\ref{alg:hardcircle} with $M=5$ ($L=32$ phases) and lies on the unit circle to numerical precision. The gap loss and log-repulsion of \eqref{eq:angle-obj} spread the phases nonuniformly over the full circle without large gaps; with cardinality matched to a uniform $2^B=32$ grid, any performance difference reflects phase \emph{locations}, not alphabet size. All experiments use this single learned instance; re-learned, uniform, or larger ($M>5$) dictionaries and joint dictionary--angle design are left to future work (Section~\ref{sec_open}).
\subsection{The order-4 simulations}
\label{subsec:order4-obstruction}
For the full physical RIS-HUBO at $N=12$, $M=5$ ($n=60$), state-vector simulation is infeasible ($2^{60}$ amplitudes), so we use the Qiskit Aer matrix-product-state (MPS) backend at bond dimension $\chi=32$, truncation $\epsilon=10^{-6}$, as in~\cite{gulbahar2026recursive}. For each of $N_I=10$ Rayleigh instances we optimize local-field $V_p$ angles for $p=1,\ldots,5$ via the provisional factorial rule \eqref{eq:cq-local-factorial-v15} and report $r_{\rm b}(i,p)=\max_{s}\|Az(s)\|_2^2/C_{\mathrm{opt}}(i)$ ($4096$ shots) against the classical reference $C_{\mathrm{opt}}(i)$. $r_{\rm b}$ is \emph{flat} at $\approx0.86$ across $p=1,\ldots,5$ and all ten channels (Fig.~\ref{fig:q2-climb}(a), robust to a $\chi{=}64$ spot-check): depth lifts the bulk of the distribution but not its extreme. The mechanism is concentration: the optimum sits many standard deviations above the random-phase mean $C_0$ with most variance in the cubic/quartic sectors.  This plateau is qualitatively consistent with the constant-level obstruction proved for pure even-$q\!\ge\!4$  models~\cite{BassoFOCS22} and with constant-depth limitations \cite{bravyi2020obstacles}; an analogous density-gap result for correlated mixed-order RIS-HUBO is an open issue.

\subsection{Order-2 QAOA with local-field $V_p$ angles}
\label{sec:q2qaoa}
We design QAOA angles for  order-2 Hamiltonian \eqref{eq:q2H} with  $V_p$ machinery of Section~\ref{sec:Vp} restricted to  quadratic sector:
\begin{equation}
c=[\,0,\,c_2,\,0,\,0\,], \,\, c_2=\sqrt{1!\,D_2\overline V_2}=0.300,\,\, q_{\max}=2 
\label{eq:q2c} 
\end{equation}
where $c_2=\sigma_J\sqrt n$, with $\sigma_J$ the per-coupling standard deviation, is the \emph{measured} per-spin local-field scale of the Frobenius-normalized Rayleigh ensemble (the $q{=}2$ case of \eqref{eq:cq-local-factorial-v15}, i.e.\ the standard deviation of $h^{(2)}_u$; $0.302$ at $N{=}16$, $0.304$ at $N{=}24$), not a tuned hyperparameter.
This physical $c_2$, not an ordered-tensor coefficient $c_2/\sqrt2$, is passed to the fixed point of Appendix~\ref{app:vp-iter}; since that recursion sums ordered pairs, the returned $V_p$ is the doubled density and the physical one is $(\langle C\rangle-C_0)/n=V_p/2$. Table~\ref{tab:q2-angles} lists solver angles at $c_{2,\mathrm{ref}}{=}0.300$; the circuit uses $\gamma_{\mathrm{circuit}}=\gamma_{V_p}\,c_{2,\mathrm{ref}}/c_2$, $\beta_{\mathrm{circuit}}=\beta_{V_p}$ (no recomputation), verbatim at the baseline $c_2{=}c_{2,\mathrm{ref}}$. This is fixed by the exact single-layer identity $V_1/2=0.091=c_2\times0.3032$~\cite{farhi2022quantum}, matching the closed-form $p{=}1$ value of Section~\ref{sec:vp_benchmark}.   Multi-start optimization of $V_p$ yields the constant-level angle pack, with $V_p$ monotone in depth.  Table~\ref{tab:q2-angles} gives the single channel-independent set $(\boldsymbol\gamma,\boldsymbol\beta)$ to $p=9$, reused verbatim across all ten channels and sizes; only the scalar $c_2$ is instance-dependent, applied through the convention above, as validated for ML MIMO detection~\cite{gulbahar2024mimo,gulbahar2025qaoa}.
\begin{table}[!t]
\centering\footnotesize
\caption{$V_p$-designed order-2 QAOA angles (full pack to $p=9$ in~\cite{repository}).}
\label{tab:q2-angles}
\setlength{\tabcolsep}{3pt}
\renewcommand{\arraystretch}{1.0}
\begin{tabular}{|c|c|ccc|ccc|}
\hline
$p$ & $V_p$ & $\gamma_1$ & $\gamma_2$ & $\gamma_3$ & $\beta_1$ & $\beta_2$ & $\beta_3$\\
\hline\hline
$1$ & $0.182$ & $1.6644$ &          &          & $0.3927$ &          & \\
$2$ & $0.245$ & $1.2752$ & $2.2104$ &          & $0.4954$ & $0.2687$ & \\
$3$ & $0.284$ & $1.0967$ & $1.8851$ & $2.1320$ & $0.5488$ & $0.3666$ & $0.2109$\\
\hline
\end{tabular}
\end{table}
\begin{figure}[!t]
\centering
\includegraphics[width=3.5in]{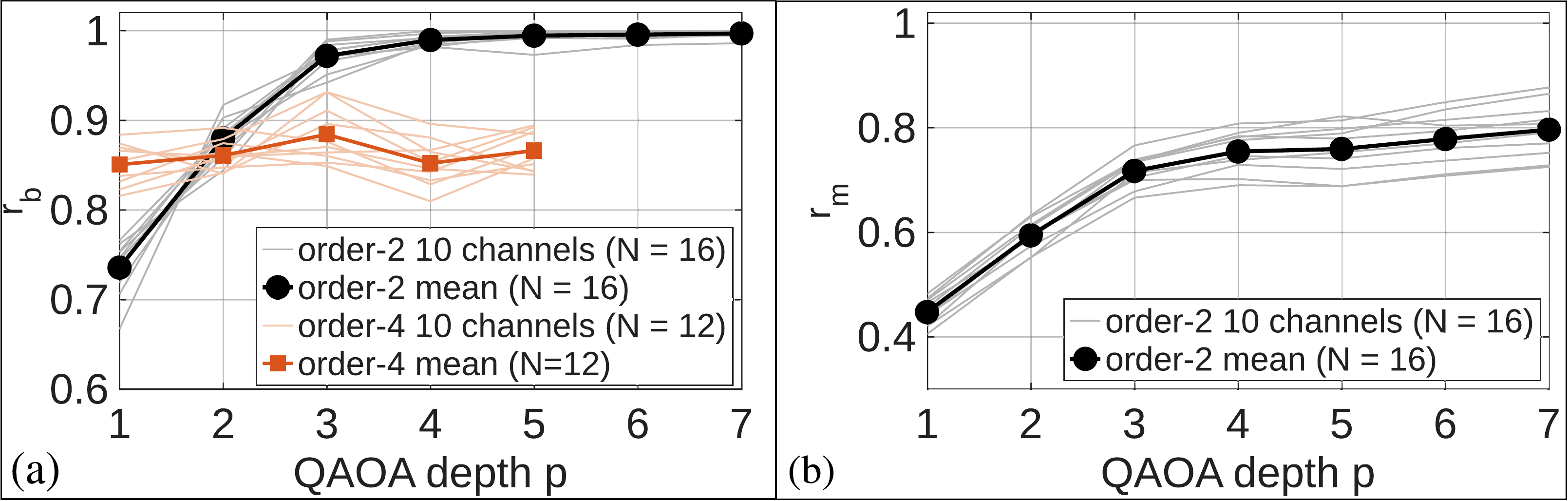}
\caption{Order-2 ($N{=}16$) and order-4 ($N{=}12$) RIS QAOA vs.\ depth $p$ for ten channel realizations and their across-channel mean. (a)~$r_b$; (b)~$r_m$.}
\label{fig:q2-climb}
\end{figure}
We simulate the full $N=16$, $M=5$ order-2 Hamiltonian ($n=80$ spins) of \eqref{eq:q2H} on the Aer MPS backend ($\chi=64$, $\epsilon=10^{-6}$) with $N_s = 4096$, showing $r_b$ against the order-2 reference $C_{\mathrm{opt}}$ averaged over the same ten channels (Fig.~\ref{fig:q2-climb}); the reference $C_{\mathrm{opt}}$ coincides with the brute-force optimum at $N=5,6$ and is the single-flip local-search value otherwise. $r_b$ climbs monotonically toward the reference (Fig.~\ref{fig:q2-climb}(a)): the mean rises $0.74\!\to\!0.88\!\to\!0.97\!\to\!0.99\!\to\!0.994$ over $p=1,\ldots,5$ and saturates at $0.996$--$0.997$ for $p=6,7$, with five of ten channels reaching the reference value $C_{\mathrm{opt}}$ and the worst at $0.986$. It is bond-converged at $p=3$: the ratio is $0.978$ ($\chi=64$) vs.\ $0.977$ ($\chi=96$, $\epsilon=10^{-7}$), the agreement of two bond dimensions supporting numerical convergence at this depth. Exact rescoring of each sampled bitstring removes cost-evaluation error, but the finite-bond sampling distribution is not a one-sided bound, and the higher-depth values are finite-bond estimates. $r_m$ (Fig.~\ref{fig:q2-climb}(b)) rises from $0.45$ ($p=1$) to $0.80$ ($p=7$) and, unlike $r_b$, keeps rising past $p=5$: $p$ shifts the distribution toward high power. This depth-improvement of the bulk matches  $V_p$ prediction of QAOA for the quadratic ensemble, which lies outside the scope of the even-$q\!\ge\!4$ obstruction theorem~\cite{BassoFOCS22}. We do not read the monotonicity itself as a no-overlap-gap certificate: a depth-$(p{+}1)$ circuit reproduces depth $p$ by zero-angle padding, so the optimized energy is non-decreasing in $p$ irrespective of any overlap gap.
\subsection{Exact scaling checks}
\label{subsec:exact-small-N}
Exact enumeration at $N=5,6$ confirms $r_b$ reaches one by $p=3$; there $r_b$ saturates trivially ($4096$ shots over $2^{25}$), so the mean is the discriminating metric, as at larger $N$. The exact single-layer ($p{=}1$) mean is known in closed form at every size; the exact small-$N$ sweep extends this to $p=2,3$ over $N=5$--$12$. Fig.~\ref{fig:psuccess}(a) shows $r_m$ (state-vector at $N=5$, MPS bond $256$ at $N=6$-$12$; $C_{\mathrm{opt}}$ exact by enumeration at $N\le6$). 
At each depth, $r_m$ decays only mildly with size ($\le\approx0.10$ across $n=25\!\to\!60$) and the depth gain is uniform ($\approx0.13$ from $p=1$ to $p=3$ at every $N$), bracketing the large-$N$ $V_p$ benchmark (Fig.~\ref{fig:ratio_vs_N}) and the $N=16$ study (Fig.~\ref{fig:q2-climb}); higher-depth/larger-size limits are discussed in Section~\ref{sec_con}.

\subsection{Instance-free $V_p$ benchmark and sampling probability} 
\label{sec:vp_benchmark}
Per-instance evaluation of mean energy is costly (PPS/MPS estimators grow heavy at $p\!\ge\!2$ for $n\gtrsim80$) but unnecessary for the \emph{mean}:    $J_{uv}$ couplings are dense and zero-mean to leading order, so $\langle C\rangle/n$ concentrates and  Sherrington--Kirkpatrick (SK) value $V_p$ in \cite{Basso2021} benchmarks against $C_{\mathrm{opt}}$ across size with no per-instance averaging, its surrogate fidelity to the finite-size RIS-HUBO being validated at $p\!\le\!2$ below.

\subsubsection{The $V_p$ map}
 $V_p$ is the SK ($q{=}2$, two-body) value, evaluated by the FWHT
fixed-point solver of Appendix~\ref{app:vp-iter} ($O(p\,4^p)$ per sweep) and
equivalently by the $O(p^2 4^p)$ iteration of~\cite{Basso2021}. With the $i{<}j$ convention used here, the fixed-point recursion returns twice
the per-spin density, so $(\langle C\rangle-C_0)/n = V_p/2$.
 
\subsubsection{Validation at $p=1$ and $p=2$}
\begin{figure}[t]\centering
\includegraphics[width=3.25in]{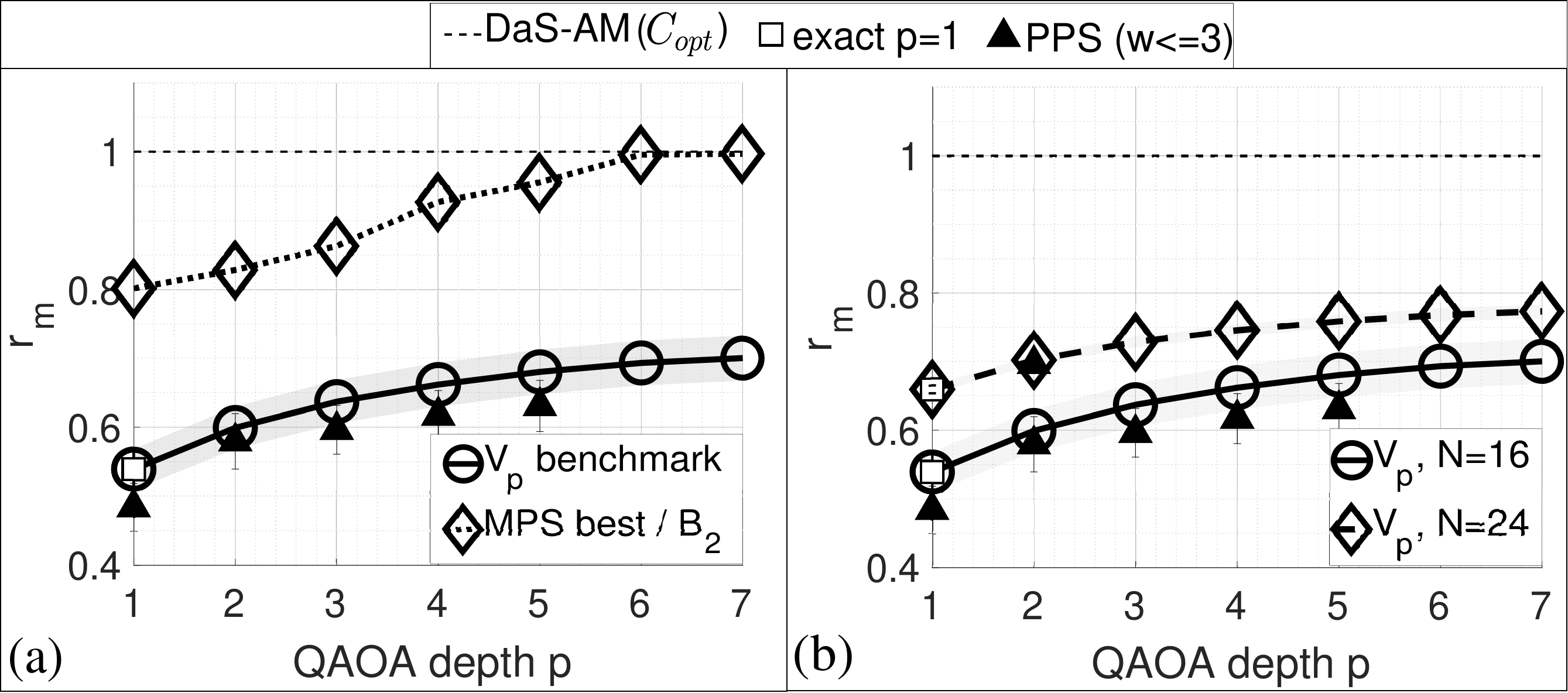}
\caption{Instance-free $V_p$ benchmark for the QAOA mean energy $r_m$ against  ($C_{\mathrm{opt}}$, dashed); the exact $p\!=\!1$ ($\square$) and
PPS $p\!=\!2$ ($\triangle$) points lie on each curve.
(a)~$N\!=\!16$ (band: instance spread); the dotted line is the best \emph{sampled}
solution from the MPS runs.
(b)~$V_p$ benchmark at two sizes.}
\label{fig:vp_N16}
\end{figure}

At $p\!=\!1$ the QAOA energy has an exact closed form for any $N$~\cite{ozaeta2022qaoa}; at $p\!=\!2$ we use PPS~\cite{angrisani2024pauli}, evolving the cost in the Heisenberg/Pauli-string picture and discarding Pauli weight above $w$ ($w\!\le\!3$). Figs. ~\ref{fig:vp_N16}(a), (b) overlay these on the $V_p$ curve, with quantitative agreement (Table~\ref{tab:vp_valid}). Although the correction is calibrated only at $p\!=\!1$, $V_p$ reproduces the independent PPS ensemble at $p\!=\!2$ to $-0.2\%$ ($N\!=\!16$) and $+1.2\%$ ($N\!=\!24$). The weight truncation is non-monotonic ($w\!=\!2$ over-shoots, $w\!=\!3$ corrects down, $w\!=\!4$ is memory-prohibitive at $n\!=\!120$);  $V_p$ cross-check confirms $w\!\le\!3$  convergence.
\begin{table}[t]\centering\small
\caption{$V_p$ benchmark versus exact/PPS data, $r_m$.}
\label{tab:vp_valid}
\setlength{\tabcolsep}{4pt}
\begin{tabular}{lcccc}
\toprule
 & $V_p$, $p{=}1$ & $V_p$, $p{=}2$ & exact $p{=}1$ & PPS $w{\le}3$, $p{=}2$\\
\midrule
$N=16$ & $0.539$ & $0.599$ & $\approx 0.539$ & $\approx 0.600$\\
$N=24$ & $0.662$ & $0.704$ & $\approx 0.662$ & $\approx 0.696$\\
\bottomrule
\end{tabular}
\end{table}

\subsubsection{Size dependence}
\begin{figure}[t]\centering
\includegraphics[width=\linewidth]{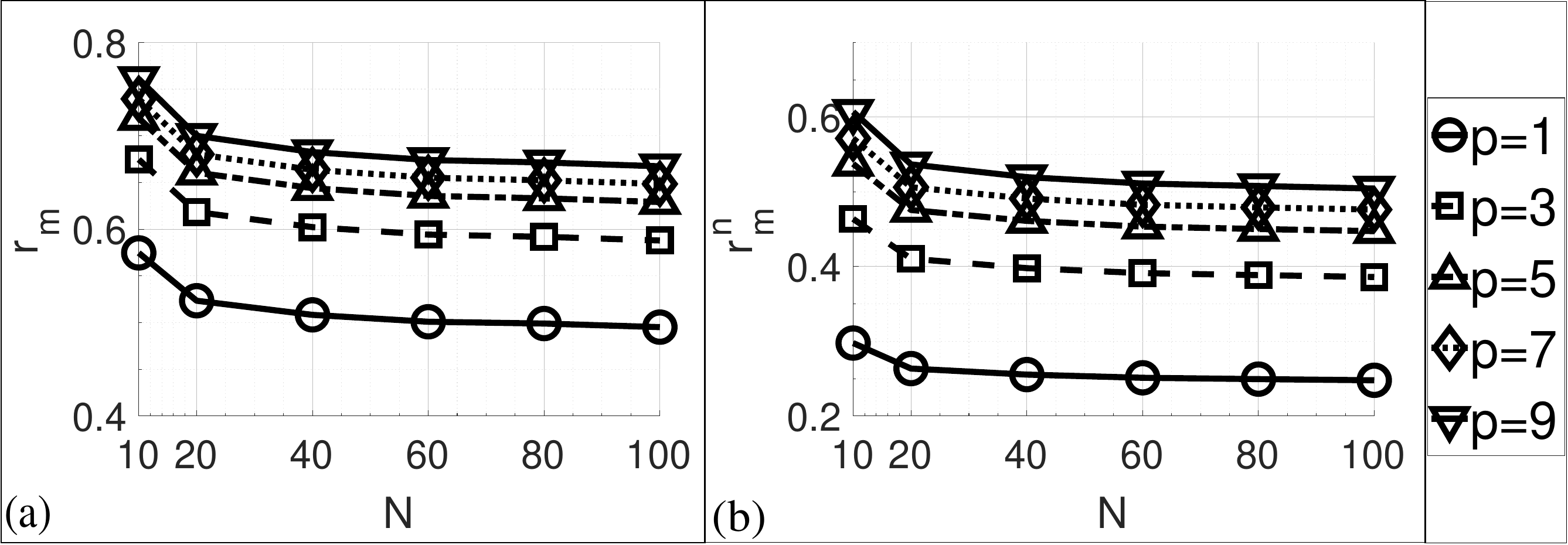}
\caption{$V_p$-normalized benchmark versus system size $N$, depths $p=1,3,5,7,9$ overlaid.
(a)~full-objective $r_m$.
(b)~offset-subtracted
$r_m^{n}$.}
\label{fig:ratio_vs_N}
\end{figure}

Fig.~\ref{fig:ratio_vs_N} plots both metrics against $N$ ($n=NM$): each rises with depth $p$ and, at fixed depth, decreases slightly with $N$ before saturating by $N\!\gtrsim\!40$. The full ratio $r_m$ (Fig.~\ref{fig:ratio_vs_N}(a)) plateaus from $\approx\!0.50$ ($p\!=\!1$) to $\approx\!0.66$ ($p\!=\!9$), but is optimistic: a large offset ($C_0\!\approx\!0.80\,\langle C\rangle$) enters both $\langle C\rangle$ and $C_{\mathrm{opt}}$ and pushes the ratio up. The more conservative metric is the offset-subtracted $r_m^{n}\equiv(\langle C\rangle-C_0)/(C_{\mathrm{opt}}-C_0)$ (Fig.~\ref{fig:ratio_vs_N}(b)), the fraction of optimal \emph{frustrated} energy captured: identical pattern (plateau $\approx\!0.27$ at $p\!=\!1$ to $\approx\!0.50$ at $p\!=\!9$, flat in $N$) at the correct lower level. We report both and base claims on the offset-subtracted one. Beyond the $p\!\le\!2$ fidelity checks of Table~\ref{tab:vp_valid}, the $p\!\ge\!3$ portion of these $V_p$ curves is the infinite-size surrogate read as a prediction, an extrapolation of the validated surrogacy, corroborated, but not independently certified, by the finite-bond $N\!=\!16$ climb (Fig.~\ref{fig:q2-climb}) and the exact small-$N$ data to $p\!=\!3$ (Fig.~\ref{fig:psuccess}).
These size-stable ratios rely on SK $V_p$ remaining faithful even though  $J$ of \eqref{eq:q2H} is far from i.i.d.: its participation-ratio effective rank is small ($r_{\mathrm{eff}}\!=\!(\sum_i\lambda_i^2)^2/\sum_i\lambda_i^4$, with $r_{\mathrm{eff}}/n\!\approx\!0.13$--$0.16$, e.g.\ $12.6/80$ at $N\!=\!16$), so the couplings are correlated rather than independent. A direct $p\!=\!1$ check across $N$ confirms the surrogate: the exact RIS energy density $d_1\!\equiv\!(\langle C\rangle-C_0)/n$, a mean/variance-matched i.i.d.\ Gaussian ensemble, and $V_p$ agree to within a few percent and stay flat in $n$ ($d_1\!\approx\!0.09$--$0.10$), with $\rho\!\equiv\!d_1^{\mathrm{RIS}}/d_1^{\mathrm{Gauss}}\!\approx\!1.05$. Low-rank correlations therefore add a small positive uplift without breaking  mean-field prediction, so $V_p$ is a quantitative instance-free surrogate for $\langle C\rangle$ at scale; any residual gap is removed by anchoring on the exact $p\!=\!1$ density, $(\langle C\rangle-C_0)/n|_p\!=\!d_1^{\mathrm{exact}}(N)\,V_p/V_1$, one $O(n^3)$ evaluation per $N$.
 
\subsubsection{Single-shot probability of an approximately optimal sample}
\label{sec:psuccess}
\begin{figure}[!t]
\centering
\includegraphics[width=\columnwidth]{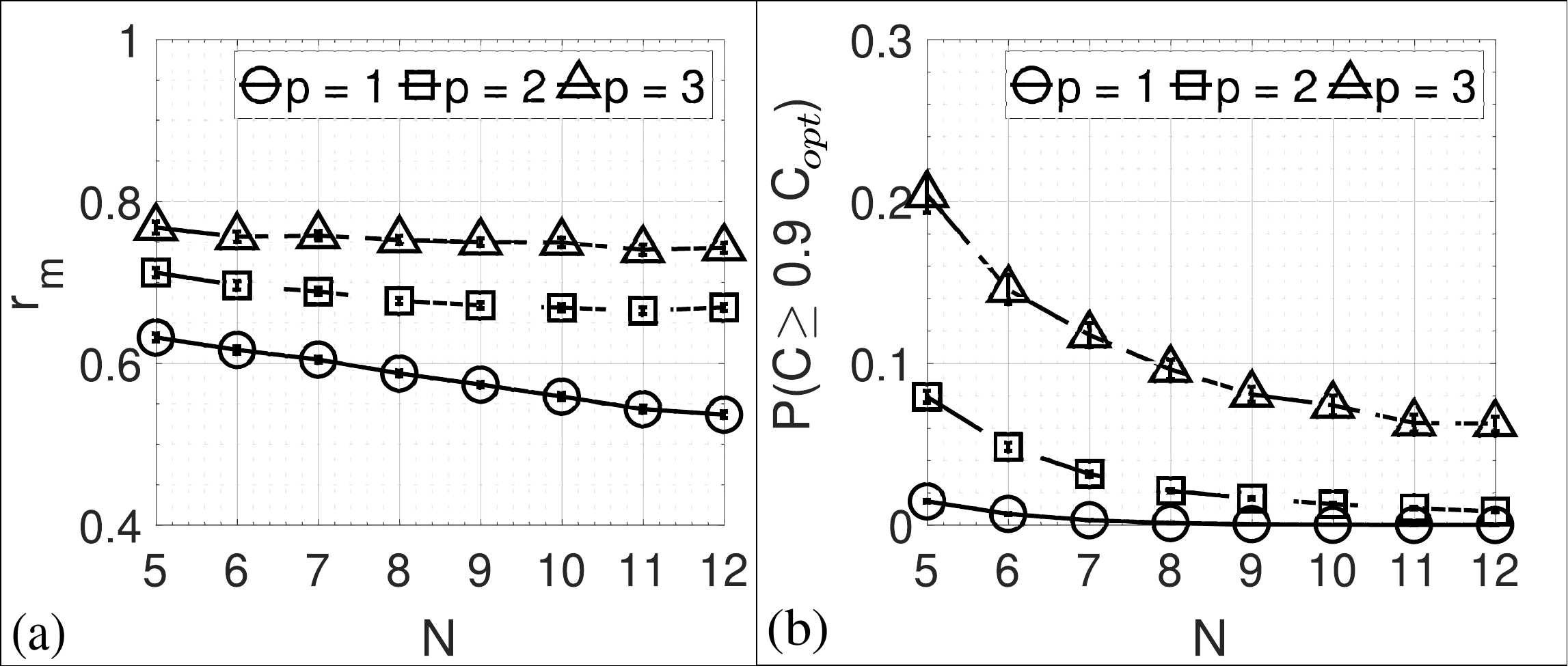}
\caption{Exact small-$N$ ensemble metrics vs.  $N$ for depths $p=1,2,3$
($V_p$-designed angles; error bars $\pm1$~SE over the instance ensemble; $100$ instances at
$N\le9$ and a uniform $50$ at $N=10$--$12$). \ (a)~the ensemble-mean ratio
$r_m$,   (b) ~the
single-shot approximate-ratio success $P(C\ge0.9\,C_{\mathrm{opt}})$ ($\alpha=0.9$).}
\label{fig:psuccess}
\end{figure}

A sharper metric is the within-$\alpha$ success probability $P_\alpha=\Pr[C(z)\ge \alpha \, C_{\mathrm{opt}}]$ that one shot lands within $\alpha$ of $C_{\mathrm{opt}}$. We estimate it from $4096$ measurements of the final state, $\alpha=0.9$, $V_p$-designed angles, on the dense order-2 instances for $N=5$-$12$ ($n=25$-$60$). Each $C(z)$ is scored exactly, so $P_{0.9}$ is unbiased up to MPS truncation. At fixed size, $P_{0.9}$ rises monotonically with depth as shown in  Fig.~\ref{fig:psuccess}(b) ($0.003\to0.032\to0.117$ at $p=1,2,3$ for $N=7$, $\sim\!40\times$ over three layers); at fixed depth, it decays only mildly with size ($P_{0.9}\!\approx\!0.12\to0.06$ from $N=7$ to $N=12$ at $p=3$, a factor of two across $n=35\to60$). At the largest depth, the order-2 QAOA returns a configuration exceeding $0.9\,C_{\mathrm{opt}}$ with $O(10^{-1})$ probability per shot at every size.
\subsection{Robustness under Rician, cascaded, and spatially-correlated RIS channels}
\label{subsec:phys-robust}
\begin{figure}[!t]
\centering
\includegraphics[width=\columnwidth]{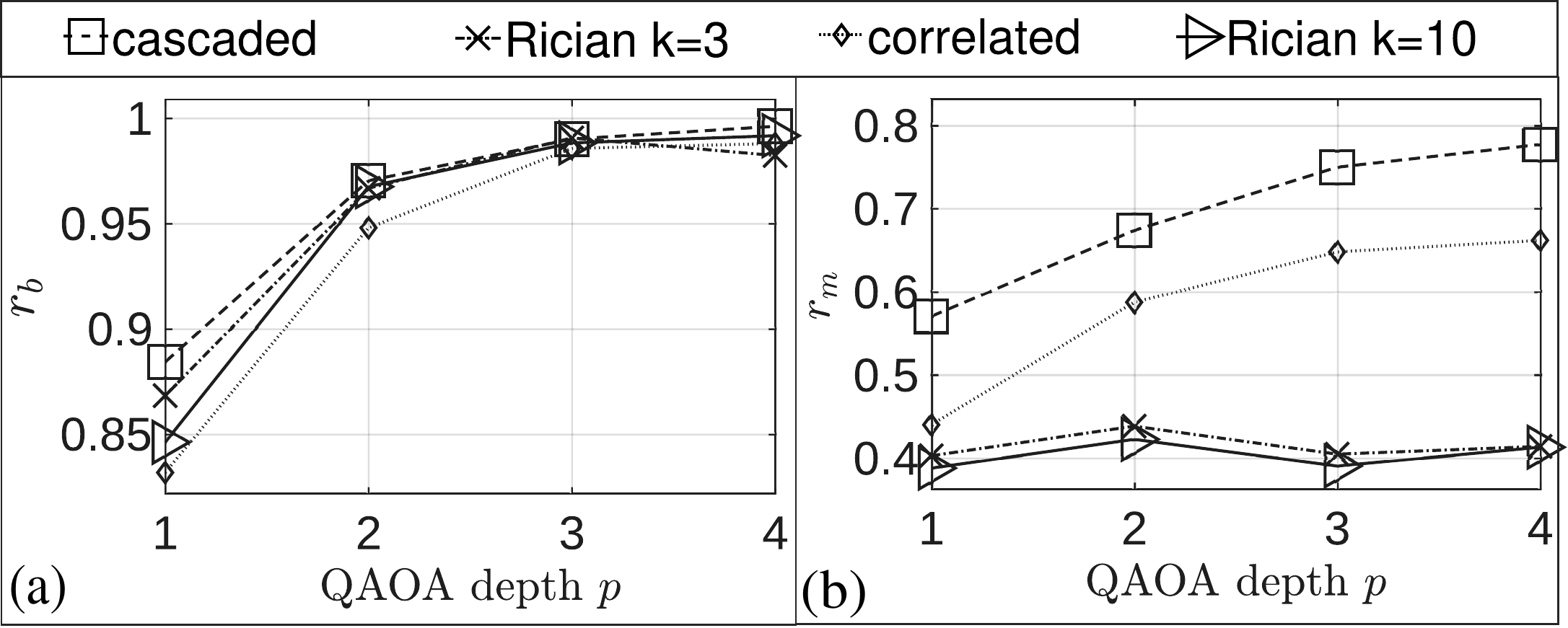}
\caption{Offline-angle transfer at $N\!=\!12$ ($n\!=\!60$ spins) on the Aer MPS backend vs. depth $p$: (a)  $r_b$ and (b) $r_m$, for the cascaded double-fading, spatially-correlated ($d\!=\!\lambda/4$), and Rician/LoS ($\kappa\!=\!3,10$~dB) ensembles.}
\label{fig:phys-transfer}
\end{figure}

\begin{figure}[!t]
\centering
\includegraphics[width=\columnwidth]{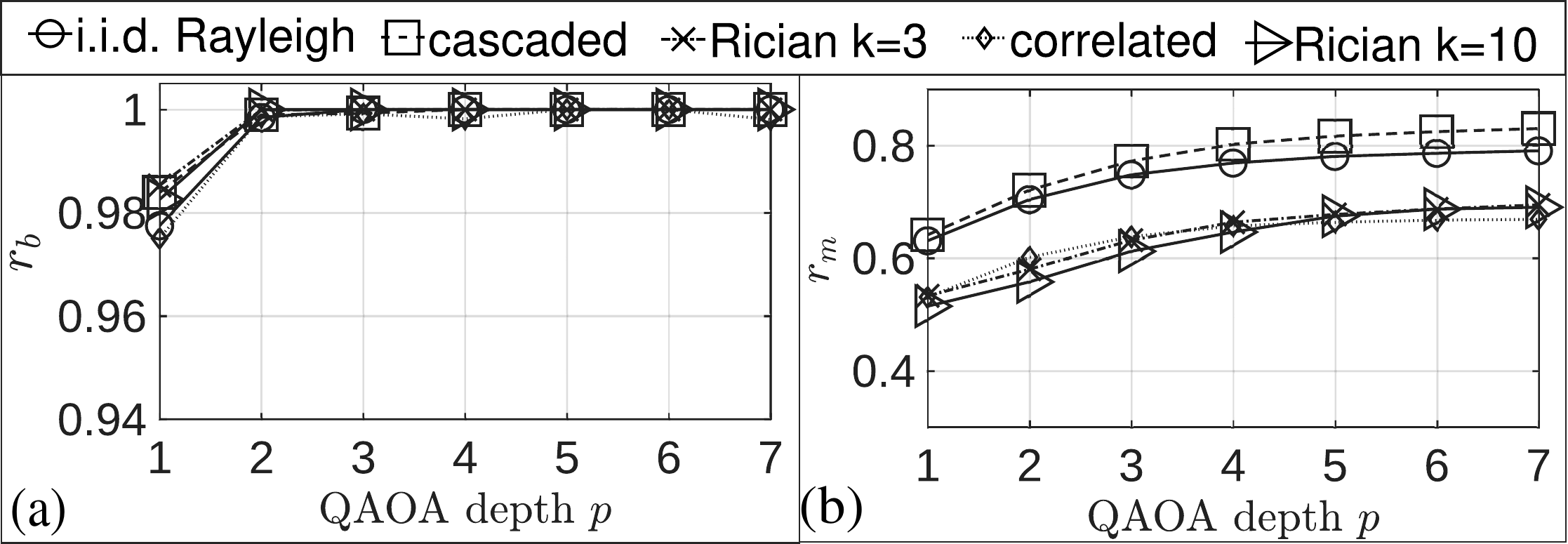}
\caption{Offline-angle transfer at $N\!=\!5$ vs. depth $p$: (a)  $r_b$ and (b) $r_m$, for the i.i.d.\ Rayleigh, cascaded double-fading, spatially-correlated ($d\!=\!\lambda/4$), and Rician/LoS ($\kappa\!=\!3,10$~dB) ensembles.}
\label{fig:phys-N5}
\end{figure}

We re-run the order-2 study on the three physical ensembles M2--M4 of Section~\ref{sec:ris-A-unified}, Rician/LoS, cascaded double-fading, and spatially-correlated fading, each built from the same $K\!\times\!N$ kernel as the i.i.d.\ baseline. The single offline angle pack of Table~\ref{tab:q2-angles} is reused; the only per-realization quantity is the scalar $c_2$ of \eqref{eq:q2c}, re-measured per channel: the rank-one LoS specular component inflates it ($\approx\!0.48,0.56$ at $\kappa\!=\!3,10$~dB), spatial correlation raises it mildly ($\approx\!0.36$), and the double-fading cascade keeps it at or below the design point ($\approx\!0.24$). Rician families need a single-scalar rescaling $\gamma_{\mathrm{circuit}}\!=\!\gamma_{V_p}\,c_{2,\mathrm{ref}}/c_2$ ($c_{2,\mathrm{ref}}{=}0.300$) to re-match the inflated scale. We evaluate at two sizes: $N\!=\!5$ ($n\!=\!25$) by exact state vector over $10$ realizations (Fig.~\ref{fig:phys-N5}) and $N\!=\!12$ ($n\!=\!60$) on the Aer MPS backend ($\chi\!=\!224$, $\epsilon\!=\!10^{-8}$, $8192$ shots, $10$ realizations; Fig.~\ref{fig:phys-transfer}). At $N\!=\!5$, $r_b$ reaches the optimum by $p\!=\!3$   and  $r_m$ rises monotonically as shown in Figs. \ref{fig:phys-N5}(a) and (b), respectively.  At $N\!=\!12$, $r_b$ climbs to within $1$--$2\%$ of the reference by $p\!=\!3$ in Fig.~\ref{fig:phys-transfer}(a), confirming the transfer for the deployed quantity. The cascaded and correlated means rise monotonically over $p\!=\!1$-$4$ (Fig.~\ref{fig:phys-transfer}(b)), whereas the Rician means stay markedly lower and are mildly non-monotone: the single-scalar rescaling matches the LoS local-field \emph{scale} but not its rank-one coupling \emph{structure}, so the structure-free offline angles improve the bulk distribution with depth for the cascaded and correlated ensembles but not for the LoS-dominated ones. The Rician $\kappa\!=\!3$ $r_b$ also eases slightly at $p\!=\!4$ ($0.991\!\to\!0.982$), which we attribute to finite-bond MPS truncation, the rank-one LoS state being the hardest to represent at fixed $\chi$; the finite-bond sample is an estimate rather than a one-sided bound. The transfer is not tied to the i.i.d.\ assumption: re-measuring $c_2$ (and, for strong LoS, the one-line rescaling) carries the channel-independent angles to LoS, cascaded, and spatially-correlated channels at $N\!\in\!\{5,12\}$, with the deployed $r_b$ reaching the reference (the certified optimum at $N\!=\!5$) across all four.
\section{Open Issues and Discussions}
\label{sec_open}
Several concrete directions remain open:
\begin{itemize}[leftmargin=1.3em,itemsep=2pt,topsep=2pt,parsep=0pt]
\item \emph{Dictionary sensitivity.} All results use one learned $L{=}32$ dictionary; re-learned, uniform, or task-adapted dictionaries and joint dictionary--angle design are untested.
\item \emph{$V_p$-to-HUBO mapping.} A theory for when correlated, non-Gaussian local fields are captured by the moments $D_q\overline V_q$ and why the factorial rule $c_q^2=(q-1)!\,D_q\overline V_q$ transfers.
\item \emph{Larger benchmarks.} Scaling the exact-benchmark check of the HUBO expansion should pair the factorial angle pack with truncated tensor networks, circuit compression, and stronger classical heuristics.
\item \emph{Hardware regime.} The deep, densely-connected order-2 circuits at large $N$ exceed NISQ coherence and connectivity budgets, so we target FTQ regime and evaluate here in simulation; a noisy NISQ study with limited connectivity (and the \textsc{swap} overhead of Section~\ref{sec:complexity}) is left open.
\end{itemize}

\emph{Classical tractability of the order-2 regime.} The order-2 reduction maps the problem to a SK quadratic spin glass, classically easy on average: for the \emph{i.i.d.\ Gaussian} SK model, message-passing reaches the optimal energy density in $O(n^2)$ time~\cite{montanari2021sk}, consistent with our spectral/single-flip reference attaining $r_b\!\to\!0.99$. That \emph{optimality} guarantee, however, assumes i.i.d.\ Gaussian couplings, whereas the order-2 coupling $J$ of \eqref{eq:q2H} (built from $G=A^{\mathsf H}A$) is low-rank and correlated, and the physical channels of Section~\ref{subsec:phys-robust} depart further still, so it does not certify the RIS problem verbatim. Classically and quantumly the picture matches: message-passing and our improving-with-depth QAOA both lie in the no-overlap-gap regime, while the order-4 sector obstructs both~\cite{BassoFOCS22}. The same offline-angle QAOA still transfers across all four physical ensembles on re-measuring only $c_2$, leaving  quantum optimization on these non-i.i.d.\ instances an open direction.

\section{Conclusions}
\label{sec_con}
We recast discrete-phase RIS power aggregation  $\|\mathbf{A} \, e^{j\theta}\|^{2}$, for a general channel $A$, as an exact order-4 HUBO with an order-2 Ising surrogate from binary sign templates. The design bounds the interaction order at four for \emph{any} $L=2^{M}$ resolution for angle set $\theta$, versus $2 \,M$ for a direct binary encoding, decoupling resolution from spin-Hamiltonian order.  That order tracks whether constant-level QAOA reaches the optimum: the order-4 sample plateaus, consistent with the obstruction proved for pure even-$q\!\ge\!4$ models, while the quadratic sector lies outside that theorem's scope, with angles fixed offline from the instance- and size-independent $V_p$ (cost $\propto 4^p$, no variational training or barren plateaus~\cite{colella2025barren}). On i.i.d.\ Rayleigh (to $N\!=\!16$, $p\!=\!7$) the best sample approximately reached the order-2 reference by $p\!=\!5$, with an exact single-layer result to $N\!=\!100$, and the same angles transfer to Rician/LoS, cascaded, and spatially-correlated channels at $N\!\in\!\{5,12\}$ on re-measuring the scalar $c_2$. QAOA-based RIS power aggregation is a candidate method for near-optimum performance with FTQ computers in the near future enabling the higher-depth circuits larger arrays need.

\section*{Acknowledgments}
The numerical calculations  were  partially performed at the TUBITAK ULAKBIM High Performance and Grid Computing Center (TRUBA) and the National Center for High Performance Computing of Turkey (UHeM) under grant number 1026432026. 
\section*{Data and Code Availability Statement}
All code and data are provided as a self-contained, pure-Python Code Ocean compute capsule~\cite{repository}:  order-2 and order-4 RIS-HUBO optimizers (hard-unit-circle dictionary synthesis, HUBO assembly, offline $V_p$ angle design,  Qiskit/Aer state-vector and MPS simulators, a Pauli-path simulator, and the classical single-flip reference) together with the scripts that regenerate every figure in the paper.

\appendices
\section{The BGMZ  Iteration and Its FWHT Evaluation}
\label{app:vp-iter}
The iteration below is the constant-level mixed-$q$ fixed-point formalism of BGMZ \cite{BassoFOCS22} (see also \cite{farhi2022quantum, Basso2021}); we restate it in our notation and our implementation evaluates it verbatim. Let $p\in\mathbb{Z}_{>0}$ and define the index set as follows:
\begin{equation}
\mathcal{A}\;=\;\bigl\{(a_1,\dots,a_p,a_{-p},\dots,a_{-1})\,:\,a_{\pm r}\in\{\pm1\}\bigr\}
\end{equation}
Given parameters $(\gamma,\beta)\in\mathbb{R}^{2p}$, define for each $a\in\mathcal{A}$ the
\emph{weights} and \emph{phases} as follows:
\begin{IEEEeqnarray}{rCl}
Q_a&=&\prod_{r=1}^{p}\!\left(\cos\beta_r\right)^{\frac{2+a_r+a_{-r}}{2}}
\left(\sin\beta_r\right)^{\frac{2-a_r-a_{-r}}{2}}
\,i^{\frac{a_{-r}-a_r}{2}}\\
\Phi_a&=&\sum_{r=1}^{p}\gamma_r\!\left(a_ra_{r+1}\cdots a_p\;-\;a_{-p}\cdots a_{-r}\right)
\end{IEEEeqnarray}
Let $c_q$ be the Hamiltonian mixture coefficients for clause sizes $q=1,\dots,q_{\max}$,
and denote by $g_q(\cdot)$ the ensemble function (e.g., Gaussian: $g_q(\lambda)=-\tfrac{1}{2}\lambda^2$)
and by $g_q'$ its derivative (Gaussian: $g_q'(\lambda)=-\lambda$), which enters \eqref{eq:Vp}.
The fixed point $\{W_a\}_{a\in\mathcal{A}}$ is defined by the following:
\begin{equation}
\label{eq:W-fp}
W_a = Q_a\exp\!\big[\textstyle\sum_{q=1}^{q_{\max}} q \, \!\!\sum_{\mathbf b\in\mathcal A^{q-1}} g_q\big(c_q\Phi_{ab_1\cdots b_{q-1}}\big)\prod_{t}W_{b_t}\big]
\end{equation}
where $ab$ is the bitwise product in $\mathcal{A}$ and $\mathbf b=(b_1,\dots,b_{q-1})\in\mathcal A^{q-1}$.
The constant-level value is then the following:
\begin{equation}
V_p \;=\;
-\sum_{q=1}^{q_{\max}} i\,c_q
\sum_{a_1,\dots,a_q\in\mathcal{A}}
g_q'\!\Big(c_q\,\Phi_{a_1\cdots a_q}\Big)\;
\prod_{t=1}^{q}W_{a_t}
\label{eq:Vp}
\end{equation}
\paragraph*{Efficient evaluation via XOR-convolution}
Let $M_p=|\mathcal{A}|=4^p$ and identify each $a\in\mathcal A=\{\pm1\}^{2p}$ with
$x\in\mathbb F_2^{2p}$ via $a=(-1)^{x}$ componentwise, so the bitwise product $ab$
corresponds to $x\oplus y$. For $f:\mathbb F_2^{2p}\to\mathbb C$ the (unnormalized)
Walsh--Hadamard transform ${\sf FWHT}\,f=\hat f$ and its inverse are:
\begin{equation}
\hat f(\omega)=\!\!\sum_{x\in\mathbb F_2^{2p}}\!\!(-1)^{\omega\cdot x}f(x),\qquad
f(x)=\frac{1}{M_p}\!\sum_{\omega\in\mathbb F_2^{2p}}\!\!(-1)^{\omega\cdot x}\hat f(\omega)
\label{eq:fwt-def}
\end{equation}
and the XOR (dyadic) convolution $(f\circledast g)(x)=\sum_{y}f(y)\,g(x\oplus y)$
satisfies $\widehat{f\circledast g}=\hat f\,\hat g$. The multi-fold sums in \eqref{eq:W-fp}--\eqref{eq:Vp} are XOR-convolution powers of $W$ (in \eqref{eq:W-fp}, convolved with the kernel $f_q$), evaluated as elementwise products in the transform domain:
\begin{equation}
W^{\circledast q}\;=\;{\sf IFWHT}\!\big(({\sf FWHT}\,W)^{\circ q}\big)
\label{eq:fwt-conv}
\end{equation}
so each fixed-point sweep costs $O(q_{\max}M_p\log M_p)=O(q_{\max}\,p\,4^p)$. Direct evaluation instead costs $O(4^{pq_{\max}})$~\cite{BassoFOCS22} ($O(16^p)$ for SK~\cite{farhi2022quantum}), or $O(p^2 4^p)$ for the pure-$q$ large-girth route~\cite{Basso2021}. To our knowledge this FWHT evaluation of the mixed-$q$ fixed point is new here: the transform--convolution identity \eqref{eq:fwt-conv} is classical, but its application to \eqref{eq:W-fp}--\eqref{eq:Vp} makes the general mixed-$q$ iteration as cheap per sweep as the pure-SK case. With $f_q(a)=g_q(c_q\Phi_a)$, $h_q(a)=g_q'(c_q\Phi_a)$ and $W\!\leftarrow\!Q$, each sweep updates $W\leftarrow Q\circ\exp\!\big(\sum_q q\,{\sf IFWHT}({\sf FWHT}(f_q)\circ({\sf FWHT}\,W)^{\circ(q-1)})\big)$ (under-relaxed) until the fixed-point residual $\|W-W^{+}\|$ is small (with $W^{+}$ the freshly computed right-hand side), giving $V_p=-\sum_q i\,c_q\,\langle h_q,\,{\sf IFWHT}(({\sf FWHT}\,W)^{\circ q})\rangle$.
\section{Operation-Count Derivations}
\label{app:complexity}
\emph{Offline.} Dictionary objective \eqref{eq:model-xy} evaluates rank-2 quadratic forms ($M\times M$) against  $L$ sign rows, so one evaluation is $O(LM^2)$, run once and independent of $N$ and channel. Each $V_p$ sweep (FWHT iteration of Appendix~\ref{app:vp-iter}) updates $|\mathcal A|=4^p$ weights by $q_{\max}$ XOR-convolutions of cost $O(p\,4^p)$ each, giving $O(q_{\max}p\,4^p)$ per sweep;  $2 \,p$-angle optimization is $N$-independent and run once per depth.
\emph{Per instance.} Forming $G=A_R^{\top}A_R+A_I^{\top}A_I$ and $S=A_I^{\top}A_R-A_R^{\top}A_I$ costs $O(KN^2)$ and order-2 couplings \eqref{eq:J2} $O(n^2)$, so HUBO assembly is $O(KN^2+n^2)$. With $G$ generically dense, cross-element couplings number $O(n^2)$, so each layer applies $O(n^2)$ $R_{ZZ}$ and $O(n)$ mixer rotations, $O(n^2p)$ gates per shot and $O(N_s n^2p)$ over $N_s$ shots (all-to-all; \textsc{swap} overhead is layout-dependent). Scoring a sample evaluates $\|Az\|_2^2$ in $O(KN)$, so $N_s$ samples cost $O(N_s KN)$.
\emph{Reference and benchmark.} Classical reference   is the multi-start single-flip local search: with an incrementally maintained local field one accepted flip costs $O(n)$, so a restart accepting $N_{\rm flip}$ flips costs $O(n^2{+}nN_{\rm flip})$ ($O(n^2)$ from forming the initial field), and $N_r$ restarts cost $O(N_r \, (n^2{+}\,n \,N_{\rm flip}))$, excluding one-off spectral and unimodular lifting initializations; no worst-case $N_{\rm flip}{=}O(n)$ bound is assumed.  DaS-AM solver~\cite{xiong2025das} is itself classical polynomial-time.  The exact finite-size $p{=}1$ RIS anchor (Section~\ref{sec_simul}) is one $O(n^3)$ closed-form evaluation per size.

\end{document}